\documentclass[10pt,journal,compsoc]{IEEEtran}
\usepackage{cite}
\usepackage{amsmath}
\usepackage{amsthm}
\usepackage{amssymb}
\usepackage{verbatim}
\usepackage{algorithm}
\usepackage{algorithmic}
\usepackage{graphicx}
\usepackage{ragged2e}
\usepackage{multirow}
\usepackage{booktabs}
\usepackage{makecell}
\usepackage{paralist}
\usepackage{color}
\usepackage[colorlinks,
 linkcolor=black,
 urlcolor=black,
 anchorcolor=black,
 citecolor=black]{hyperref}
\usepackage{xcolor,colortbl}
\usepackage{float}
\usepackage{epstopdf}
\usepackage{subfigure}

\floatname{algorithm}{Algorithm}

\newcommand{\SUB}[1]{\ENSURE \hspace{-0.15in} \textbf{#1}}

\begin{document}
\bstctlcite{BSTcontrol}

\title{Secure and Efficient Federated Learning Through Layering and Sharding Blockchain}

\author{Shuo~Yuan,~\IEEEmembership{Member,~IEEE,}
	Bin~Cao,~\IEEEmembership{Senior~Member,~IEEE,}
	Yao~Sun,~\IEEEmembership{Senior~Member,~IEEE,}
	Zhiguo~Wan,~\IEEEmembership{Member,~IEEE,}
	and~Mugen~Peng,~\IEEEmembership{Fellow,~IEEE}
	\IEEEcompsocitemizethanks{
		\IEEEcompsocthanksitem {This work was supported in part by the National Key R$\&$D Program of China under Grant 2021YFB1714100, in part by the National Natural Science Foundation of China under Grant U22B2006, and in part by the BUPT Excellent Ph.D. Students Foundation under Grant CX2020106.
		This work was presented in part at 2021 IEEE WCNC \cite{yuan2021chainsflblockchaindriven}.
		({Corresponding author: Bin Cao})}
		\IEEEcompsocthanksitem {Shuo Yuan, Bin Cao, and Mugen Peng are with the State Key Laboratory of Networking and Switching Technology, Beijing University of Posts and Telecommunications, Beijing 100876, China (e-mail: yuanshuo@bupt.edu.cn; caobin@bupt.edu.cn; pmg@bupt.edu.cn).}
		\IEEEcompsocthanksitem{Yao Sun is with James Watt School of Engineering, University of Glasgow, G12 8QQ Glasgow, Scotland, UK (e-mail: Yao.Sun@glasgow.ac.uk).}
		\IEEEcompsocthanksitem{Zhiguo Wan is with Zhejiang Lab, Hangzhou 311121, Zhejiang, China (e-mail: wanzhiguo@zhejianglab.com).}}
}

\IEEEtitleabstractindextext{
	\begin{abstract}
		\justifying
		Introducing blockchain into Federated Learning (FL) to build a trusted edge computing environment for transmission and learning has attracted widespread attention as a new decentralized learning pattern.
		However, traditional consensus mechanisms and architectures of blockchain systems face significant challenges in handling large-scale FL tasks, especially on Internet of Things (IoT) devices, due to their substantial resource consumption, limited transaction throughput, and complex communication requirements.
		To address these challenges, this paper proposes ChainFL, a novel two-layer blockchain-driven FL system.
		It splits the IoT network into multiple shards within the subchain layer, effectively reducing the scale of information exchange, and employs a Direct Acyclic Graph (DAG)-based mainchain as the mainchain layer, enabling parallel and asynchronous cross-shard validation.
		Furthermore, the FL procedure is customized to integrate deeply with blockchain technology, and a modified DAG consensus mechanism is designed to mitigate distortion caused by abnormal models.
		To provide a proof-of-concept implementation and evaluation, multiple subchains based on Hyperledger Fabric and a self-developed DAG-based mainchain are deployed.
		Extensive experiments demonstrate that ChainFL significantly surpasses conventional FL systems, showing up to a 14\% improvement in training efficiency and a threefold increase in robustness.
	\end{abstract}

}

\maketitle
\IEEEdisplaynontitleabstractindextext
\IEEEpeerreviewmaketitle

\section{Introduction}

\IEEEPARstart{W}{ith} the advent of the Internet of Everything era, the enormous volume of data generated by various connected devices, such as mobile phones, vehicles, and smart sensors, has become an invaluable resource for societal advancement. 
Machine Learning (ML), widely recognized for its potency and effectiveness, plays a pivotal role in utilizing this data resource, driving a variety of smart Internet of Things (IoT) applications, such as smart grids, intelligent transportation systems, and smart industries \cite{khan2021federatedlearning}. 
However, the common centralized ML methodologies, which involve collecting data from IoT devices at a central location for training, present some drawbacks. 
This centralized approach not only leads to increased transmission delays and extended learning convergence times but also poses serious concerns regarding privacy breaches and the potential for data misuse, highlighting the need for more secure and efficient data processing methods in IoT networks.

To this end, Federated Learning (FL) \cite{mcmahan2017communicationefficientlearning} as a promising training paradigm has been proposed to allow devices to collaborate and train a shared ML model in a distributed manner while keeping training data local.
The main benefit of FL is that only local models without any raw data need to be shared during the entire learning process \cite{mcmahan2017communicationefficientlearning}.
This allows FL to take full advantage of the resources and data of IoT devices to implement intelligence endogenous IoT services, such as predictive maintenance of industrial devices, traffic prediction in Internet-of-vehicle networks, and disease diagnosis based on wearable devices.
However, despite the potential benefits of traditional FL, there are several security and efficiency issues in a trustless edge computing environment that have yet to be fully addressed in practical applications, which can be summarized as follows.

\textbf{Security Issues:}
Traditional FL systems depend on a central aggregator for training orchestration, but this centralization presents security risks, including a Single Point of Failure (SPOF) and vulnerability to targeted attacks, potentially causing service disruption and paralysis \cite{cao2023ondevicefederated}. 
In addition, the potential bias from the central aggregator in selecting IoT devices each round can adversely affect global model accuracy \cite{zhao2022dynamicreweighting}. 
Moreover, traditional FL lacks mechanisms to address trust issues in the transmission and learning process of models, such as the poisonous models generated by malicious IoT devices \cite{zhu2023blockchainempoweredfederated}.

\textbf{Efficiency Issues:}
Most FL systems operate in a synchronous manner, wherein the central server waits for all participating IoT devices to submit their local models before updating the global model.
This approach is slowed down by \emph{stragglers}, devices that take longer to complete training iterations, which affect overall convergence speed \cite{dutta2018slowstale}.
In contrast, asynchronous training approaches \cite{xie2019asynchronousfederated} update the global model with potentially outdated local models, known as \emph{stale models}, which can lead to instability of the global model during the updating process.

To address the aforementioned issues, a series of works have introduced blockchain \cite{nakamoto2008bitcoinpeertopeer} into FL to exploit the advanced features of blockchain, such as tamper-resistant, decentralized, and traceability, in the trustless edge computing environment \cite{kim2020blockchainedondevice,lu2020blockchainfederated}.
In \cite{kim2020blockchainedondevice}, BlockFL is proposed to carry out synchronous FL training in a decentralized manner.
Then, the SPOF and targeted attacks can be overcome, and all local model updates are verified by blockchain nodes on the Proof-of-Work (PoW) consensus \cite{nakamoto2008bitcoinpeertopeer}.
To alleviate the computation consumption during the consensus, a collaborative system for industrial IoT is proposed in \cite{lu2020blockchainfederated} to integrate the federated training into the consensus process.
Besides, some works also introduced differential privacy into blockchain-based FL to further enhance the data privacy of IoT devices \cite{lu2020blockchainfederated}.
However, these studies fail to account for the constraint imposed by blockchain throughput, a pivotal determinant of training process efficiency.

Despite some enhancements to distributed training from blockchain-enabled systems, integrating blockchain with FL still faces notable challenges, which can be summarized as follows:
\begin{enumerate}[1)]
	\item \emph{High Computation Cost.}
	      The use of PoW or PoW-based consensus mechanisms entails solving computational puzzles to establish the right to generate blocks, ensuring blockchain stability and security.
	      However, this computational requirement introduces a significant computation cost \cite{cao2020performanceanalysis}.
	      Moreover, the time spent on solving these puzzles unavoidably slows down the convergence of the training task.
	\item \emph{Limited Scalability.}
	      As we all know, most consensus hardly handles high scalability and decentralization at the same time due to the cost of computation, communication, and time \cite{croman2016scalingdecentralized}.
	      For example, PoW consensus suffers from low transaction throughput due to intensive hash computation and cannot scale out its transaction processing efficiency with the increase of blockchain nodes.
	      Besides, Practical Byzantine Fault Tolerance (PBFT) \cite{castro1999practicalbyzantine} faces limitations imposed by network bandwidth due to the frequent communication exchanges it requires.
	\item \emph{Huge Storage Requirement.}
	      Blockchain operates as a distributed ledger, which necessitates each blockchain node to maintain a record of verified blocks in its local ledger.
	      Consequently, the limited storage capacity of nodes significantly hinders information exchange speed within the network, thereby impacting the delivery of services supported by the blockchain.
	\item \emph{Stragglers.}
	      Most of the blockchain-enabled FL systems, such as BlockFL \cite{kim2020blockchainedondevice}, PIRATE \cite{zhou2020pirateblockchainbased}, and DeepChain \cite{weng2021deepchainauditable} are processed in a synchronous manner.
	      Hence, the presence of stragglers hampers training efficiency, similar to traditional FL scenarios.
	      Currently, there is limited research on asynchronous training based on blockchain, let alone the detection of stale models.
\end{enumerate}

To address these challenges, we propose a hierarchical blockchain-driven FL system, named ChainFL, to enhance the scalability and security of decentralized FL.
By splitting the large-scale IoT network into multiple shards, the majority of information exchange and storage is limited in the same shard which significantly reduces the communication rounds and storage requirements.
Besides, the model trained from each shard can be obtained and validated by other shards efficiently with the help of the Direct Acyclic Graph (DAG) consensus-based mainchain.

Overall, the main contributions of this paper are summarized as follows:
\begin{itemize}
	\item 
	We propose ChainFL, a novel FL system driven by the hierarchical blockchain, with the aim to provide a secure and effective FL solution for large-scale IoT networks.
	We design a Raft-based blockchain sharding architecture to improve scalability and an adapted DAG-based mainchain to achieve cross-shard interactions. 
	To our knowledge, ChainFL is the first system to leverage a DAG for coordinating multiple shard blockchain networks, thereby improving the security and scalability of FL systems.
	\item 
	We define the operation process and interaction rules for ChainFL to perform the FL tasks. 
	To improve learning efficiency, synchronous and asynchronous training are combined in ChainFL to alleviate the negative impact of stragglers. 
	Moreover, a virtual pruning mechanism is designed based on the adapted DAG consensus to eliminate the impact of abnormal models.
	\item 
	We devise a sharding network prototype leveraging Hyperledger Fabric to instantiate the subchain layer of ChainFL, while developing a DAG-based blockchain to implement the mainchain layer, thereby fulfilling cross-layer interactions.
	The off-chain storage scheme is adopted in the prototype to reduce the storage requirements of blockchain nodes in both layers.
	The extensive evaluation results show that ChainFL provides acceptable and sometimes better convergence rates (by up to 14\%) compared to FedAvg \cite{mcmahan2017communicationefficientlearning} and AsynFL \cite{xie2019asynchronousfederated} for CNNs and RNNs, and enhances the robustness (by up to 3 times) of FL system.
\end{itemize}

The remainder of this paper is structured as follows.
The related works are reviewed in Section \ref{sec:related}.
Section \ref{sec:architecture} introduces the architecture of ChainFL, with Section \ref{sec:designDetails} detailing its workflow and consensus. 
Implementation and evaluations are in Section \ref{sec:implementation} and Section \ref{sec:evaluation}, respectively.
Finally, conclusions and future work are presented in Section \ref{sec:conclusion}.

\begin{figure}[H]
	\centering
	\subfigure[]{
		\label{fig:TypicalFLArchitecture}
		\includegraphics[width=0.2\textwidth]{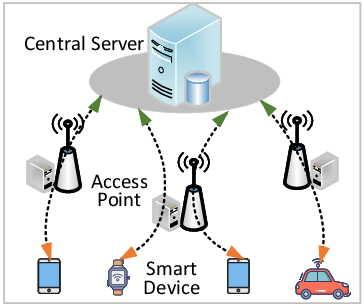}}
	\hspace{8mm}
	\subfigure[]{
		\label{fig:TypicalBlockchainedFLArchitecture}
		\includegraphics[width=0.2\textwidth]{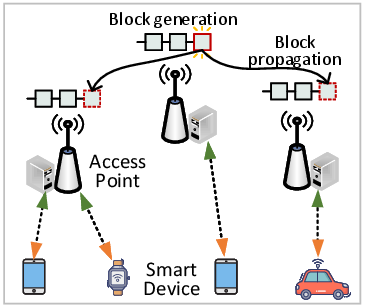}}
	\caption{Typical architecture of FL. (a) Typical master/slave architecture of FL with a central server. (b) Typical decentralized architecture of blockchained FL.}
	\label{fig:TypicalArchitecture}
\end{figure}

\section{Related Works}
\label{sec:related}

Recent studies emphasize the crucial role of blockchain in enhancing security and availability in FL.
This section reviews these works, focusing on blockchain-enabled FL, and summarizes current advancements to underscore the novelty of our work.

\subsection{Blockchain-enabled FL Framework}

As shown in Fig. \ref{fig:TypicalFLArchitecture}, traditional FL runs in a master/slave manner where the capacity and concurrency of the centralized master server handling massive participants are usually the bottlenecks to performing the distributed learning.
Recently, some works such as \cite{kim2020blockchainedondevice, qu2022blockchainenabledfederated,zhou2020pirateblockchainbased,shayan2021biscottiblockchain, zhang2021refinerreliable,majeed2019flchainfederated,liu2021blockchainfederated, jin2021crossclusterfederated, weng2021deepchainauditable} have achieved decentralized learning by using blockchain to tackle the bottlenecks, and the typical architecture is shown in Fig. \ref{fig:TypicalBlockchainedFLArchitecture} where the training process is orchestrated by distributed nodes instead of the master server.
To prevent the effect of local malicious gradients on the convergence of the global model, two Byzantine-resilient FL architectures based on blockchain, namely PIRATE and Biscotti, were proposed by the authors of \cite{zhou2020pirateblockchainbased} and \cite{shayan2021biscottiblockchain}, respectively.
Integrating edge computing in BlockFL, Majeed \emph{et al}. \cite{majeed2019flchainfederated} proposed FLchain to train multiple global models in parallel based on the channel feature of Hyperledger Fabric, Liu \emph{et al}. \cite{liu2021blockchainfederated} proposed a decentralized intrusion detection system in response to the increased cyber-intrusions in vehicular networks.
To decrease the high communication latency between devices that are separated by great distance, Jin \emph{et al}. \cite{jin2021crossclusterfederated} proposed a cross-cluster FL based on blockchain for the Internet of medical things.
To promote participant engagement and incentivize correct behavior, DeepChain \cite{weng2021deepchainauditable} and Refiner \cite{zhang2021refinerreliable} have introduced incentive mechanisms into their respective blockchain-based FL systems.
However, these works have not considered the limitation of blockchain throughput, which is a critical factor in determining the efficiency of the training process.

On the other hand, several studies have implemented blockchain protocols on devices with constrained computational and storage capacities, such as mobile phones and vehicles \cite{zhou2020pirateblockchainbased}.
These devices are tasked with updating their local models, as well as collecting other updated models and solving computationally intensive puzzles to generate blocks.
Such requirements pose significant challenges for edge devices \cite{cao2023blockchainsystems}.
Furthermore, the necessity to maintain an ever-expanding local distributed ledger can compound these difficulties, potentially diminishing the efficiency of FL.
A partial solution involves nodes retaining only a segment of the ledger to reduce storage demands. 
However, this approach necessitates frequent interactions with other entities to access information not stored locally, subsequently increasing communication overhead.

\subsection{Blockchain Consensus}
The consensus mechanism, a cornerstone of blockchain technology, is instrumental in facilitating agreement in decentralized settings \cite{cao2023blockchainsystems}. 
The PoW protocol, notably used in Bitcoin \cite{nakamoto2008bitcoinpeertopeer} and later adopted in BlockFL \cite{kim2020blockchainedondevice}, supports an open participation model without the need for authorization. 
However, PoW and similar protocols are resource-intensive and time-consuming due to their reliance on solving complex hash problems to compete for block generation rights.
In response, Proof of Federated Learning (PoFL) \cite{qu2021prooffederated} has been proposed to repurpose the computational efforts of PoW towards FL model training. 
Nevertheless, PoFL is still hampered by limited throughput and an increased likelihood of forking in scalable, competition-based consensus environments \cite{zhou2020pirateblockchainbased}.
Alternatively, the PBFT consensus mechanism requires multiple communication rounds to achieve consensus, which leads to exponentially increased communication overhead as participant numbers rise \cite{castro1999practicalbyzantine}.
In contrast, the Raft consensus \cite{ongaro2014searchunderstandable}, which relies on leader selection and log replication to achieve rapid and reliable consensus without the high computational costs and extended confirmation times associated with other protocols, resulting in a significant improvement in throughput.
However, the throughput of Raft is constrained by the peak performance of the single node with limited resources \cite{huang2020performanceanalysis}.

\subsection{Synchronous \& Asynchronous FL}
The FL process can be categorized into two types: synchronous FL and asynchronous FL.
In synchronous FL, as detailed in \cite{mcmahan2017communicationefficientlearning, qu2022contextawareonline}, training is concurrently executed by participants, with the FL aggregator awaiting the completion of all local model updates. 
The performance of synchronous FL is enhanced by optimizing participant selection in constrained wireless networks \cite{qu2022contextawareonline} and by introducing an edge device clustering and a cosine similarity-based model filter to reduce parameter exchange redundancy \cite{wang2022edgebasedcommunication}. 
However, the network on the master aggregator side can experience congestion when an excessive number of participants check in concurrently \cite{xie2019asynchronousfederated}.
In addition, the duration per iteration is likely to increase as the participant count rises, a situation exacerbated by the occurrence of straggling participants, known as ``stragglers'', who intermittently slow down the training process \cite{dutta2018slowstale}.

In contrast to synchronous FL, asynchronous FL \cite{xie2019asynchronousfederated} presents an alternative method that effectively addresses the straggler issue.
This approach involves the central server promptly updating the global model upon receiving individual local models, rather than waiting for all participant updates.
However, a potential downside of this method, as indicated in \cite{xie2019asynchronousfederated}, is the risk of destabilizing the global model due to the aggregation of stale models, which were trained based on a previous version of the global model. 
To address these issues of model staleness and to improve efficiency, the work \cite{wu2021safasemiasynchronous} has proposed a semi-asynchronous FL approach, innovating in the areas of client selection and the rules for global model aggregation. 
However, it is noteworthy that these methodologies do not take into consideration the potential impact of malicious participants, whose actions could significantly compromise the accuracy of the global model.

Although both FL and blockchain are operational in distributed networks, it is still a challenge to refine the training process to adapt to the blockchain network while effectively reducing the impact of straggler and/or stale models.

\subsection{The Novelty of the Paper}

In this paper, we consider a classic blockchain-driven distributed learning scenario, which includes devices eager to leverage their data for participation in decentralized learning processes and a large blockchain network supported by edge nodes with abundant storage and computing resources.
We exploit the sharding architecture \cite{luu2016securesharding} to split the large-scale blockchain network into multiple shards to enhance the parallelism of consensus, which significantly scales up the overall throughput and reduces the storage requirement of blockchain nodes.
Further, we design a DAG-based mainchain to enable the asynchronous processing of models trained by each shard, which can efficiently speed up the validation and aggregation of shard models.

\section{Our Proposed ChainFL System}
\label{sec:architecture}

\begin{figure}[t]
	\centering
	\includegraphics[width=0.48\textwidth]{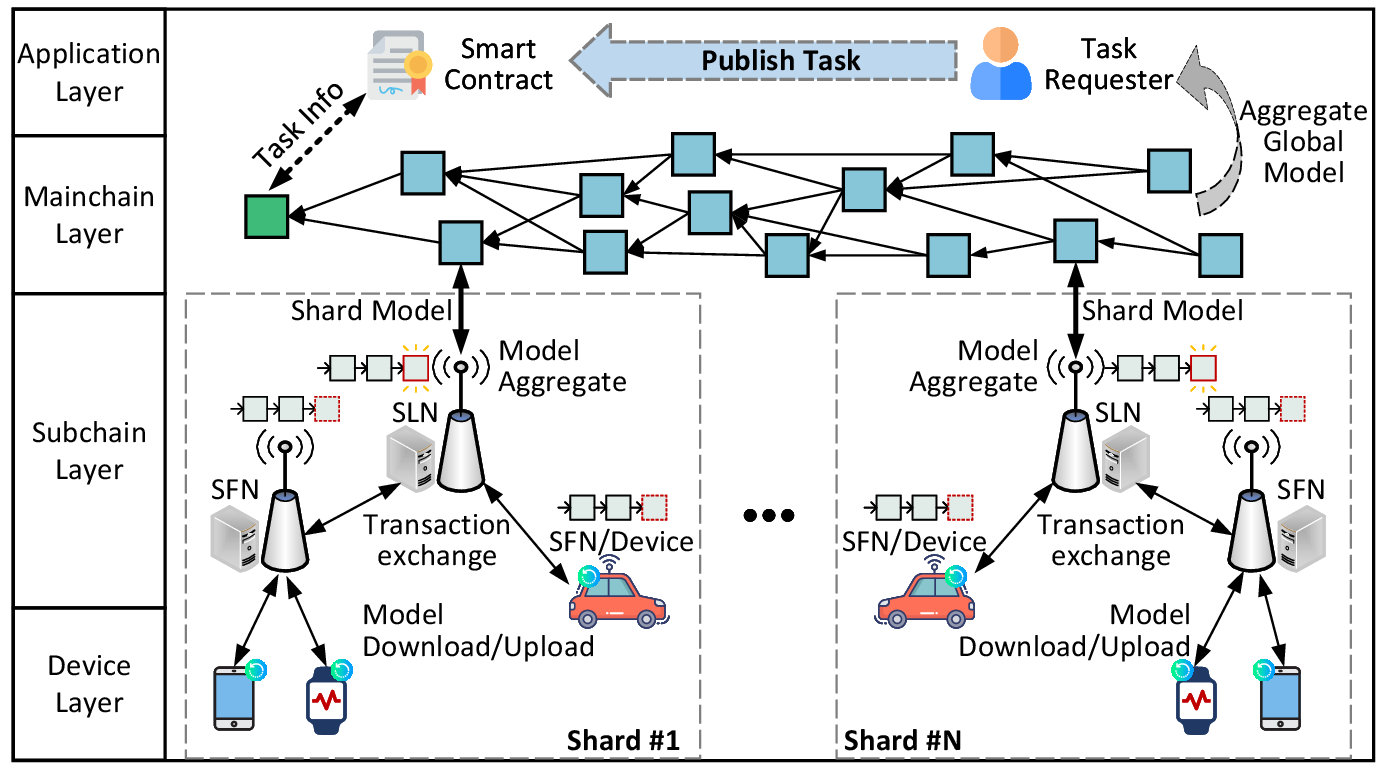}
	\caption{Layered architecture of ChainFL.}
	\label{fig:archiChainFL}
\end{figure}

In this section, we present the architecture of ChainFL and describe its main components.
As depicted in Fig. \ref{fig:archiChainFL}, ChainFL employs a two-layer blockchain architecture, which comprises a subchain layer, consisting of multiple subchains, and a mainchain layer, featuring a single DAG-based mainchain.

The subchain layer of ChainFL is based on the classic multi-access edge computing scenario \cite{cao2019intelligentoffloading} which is commonly employed in smart IoT environments to support IoT devices with limited resources.
In this scenario, edge nodes (e.g., IoT devices and access points with abundant computation resources) are partitioned into multiple independent groups (referred to as shards) that deploy subchains and act as blockchain nodes to facilitate information exchange and consensus formation.
To meet the requirements of access control for IoT devices, the consortium blockchain is adopted in the subchain.
On the other hand, the mainchain can be deployed on many distributed edge nodes or trusted computation platforms to maintain and validate transactions submitted by shards in a decentralized manner.

\emph{Shard Entities and Training Manner:}
Within each shard, there exist several entity types, including IoT devices, Subchain Leader Nodes (SLNs), and Subchain Follower Nodes (SFNs).
Each shard conducts the training task in a synchronous manner, while interactions between subchains and the mainchain occur independently and asynchronously.
As a result, ChainFL integrates both synchronous and asynchronous training manners.
Further details of the interaction process are described in Section \ref{subsec:procedure}.

To well elaborate, the layered architecture of ChainFL, as shown in Fig. \ref{fig:archiChainFL}, is described as follows.

\vspace{-0.1in}
\subsection{Device Layer}
This layer is composed of IoT devices that participate in FL tasks, such as phones, vehicles, and smart home appliances.
These devices are responsible for maintaining the collected data and training the local model.
In addition, IoT devices must pack their updated local models into transactions along with additional information, such as authorization details and timestamps, and then submit the transactions to the subchain.

\vspace{-0.1in}
\subsection{Subchain Layer}
In each shard, independent subchains are deployed, each bearing the responsibility of orchestrating the IoT devices within the shard to collaboratively accomplish the training task in a synchronous manner.
The Raft consensus \cite{ongaro2014searchunderstandable} is adopted in each subchain, and the details about this consensus in ChainFL are given in Section \ref{subsec:raftConsensus}.
In addition, the edge nodes as blockchain nodes in each subchain fall into two categories:
\begin{itemize}
	\item \emph{Subchain Leader Node (SLN).}
		  The selection mechanism for an SLN within each subchain follows the consensus protocol specific to that subchain.
	      Notably, in a Raft-based subchain, the election of an SLN is conducted via a democratic voting process, as detailed in \cite{ongaro2014searchunderstandable}.
		  Beyond executing fundamental consensus operations, the duties of the SLN include the selection of devices for participation in the training task, as well as the authorization of their access to the subchain.
		  Further responsibilities include the aggregation of local models and the uploading of the updated shard model to the mainchain upon the completion of each iteration.
		  Concurrently, the SLN constructs new basic iteration models from the mainchain for subsequent training iterations.
	\item \emph{Subchain Follower Node (SFN).}
		  Each SFN is responsible for authenticating and verifying the accuracy of the transactions (local models) before they are transmitted to the SLN. 
		  Moreover, all SFNs within a particular shard need to reach a consensus on the block generated by the SLN. 
		  This consensus is reached through adherence to the specific consensus protocol employed within that shard.
\end{itemize}

\emph{Subchain Consensus:}
To adapt to IoT scenarios and alleviate the computational burden on IoT devices, the Raft protocol, which has low computational complexity, is introduced in this paper as the consensus mechanism for each subchain\footnote{ChainFL is also able to employ alternative consensus mechanisms, such as PBFT and Proof of Stake, as long as a robust leader selection mechanism has been designed well for each shard.}.
Importantly, the inherent bottleneck of Raft (that is, the throughput limited by the performance of a single node) is effectively addressed by reducing the amount of transaction processing of the leader through sharding.
It is worth noting that IoT devices with abundant resources can participate in FL tasks, not only to train the local model but also to serve as edge nodes, thereby establishing the consensus of the subchain simultaneously.

\vspace{-0.1in}
\subsection{Mainchain Layer}
The mainchain within the proposed architecture employs an asynchronous consensus mechanism based on DAG architecture, commonly referred to as DAG consensus \cite{li2020directacyclic} or tangle consensus \cite{popov2018tangle}.
The performance and security aspects of this DAG-based mainchain have been extensively analyzed in our previous studies \cite{cao2020performanceanalysis} and \cite{li2020directacyclic}. 
As illustrated in Fig. \ref{fig:archiChainFL}, the DAG-based mainchain is characterized by \emph{vertices indicating individual transactions and directed edges representing the validation of one transaction by another}.
In this structure, each transaction encapsulates a model trained by an individual shard. 
Transactions without validation from others are termed as ``\emph{tips}.''
Distinctively, the mainchain in this system diverges from traditional PoW-based blockchain architectures by not depending on a linear chain for validation, owing to its graph-based nature. 
This inherent ability to accommodate forks enables the mainchain to process transactions asynchronously. 
Consequently, ChainFL, implemented on IoT networks, demonstrates scalable capabilities without significantly impeding system throughput. 
For the expansion of ChainFL, new IoT devices can integrate into an existing shard or collaborate with other edge nodes/devices to form a new shard. 
Furthermore, each node/platform in the mainchain network possesses a local ledger, which facilitates the construction of a DAG. 
Notably, the deployment of both the subchain and the mainchain is feasible on a single edge node, provided that the node has adequate resources.

\vspace{-0.1in}
\subsection{Application Layer}
The application layer is above the mainchain layer and utilizes the interface offered by the mainchain layer to trigger FL tasks through smart contracts\footnote{The smart contract is a self-executing contract with the terms of negotiations between users being directly written into a computer program \cite{du2023blockchainaidededge}.}.
The FL task requester publishes the task by signing a smart contract that declares its task requirements and conditions for completing the task.
Correspondingly, IoT devices and edge nodes engaged in these tasks are incentivized with specific rewards upon successful task completion.

\vspace{-0.1in}
\section{ChainFL Workflow \& Consensus}
\label{sec:designDetails}

In this section, we introduce the blockchain-enabled FL algorithm, detail the FL process and consensus mechanisms of ChainFL, and analyze the probability of tip selection.

\subsection{FL Algorithm}
As mentioned above, both synchronous and asynchronous training are integrated into ChainFL and run in a decentralized manner.
Therefore, the FL algorithm originally proposed in \cite{mcmahan2017communicationefficientlearning} needs to be modified to adapt to the architecture of ChainFL.

To describe the algorithm clearly, we take shard \#1 as an example.
We assume that the set of IoT devices $\{ {{d_1},{d_2}, \cdots,{d_n}} \}$ is selected by SLN of shard \#1 to participate in the FL task, and the datasets of these devices are $\{ {D_1},{D_2}, \cdots,{D_n}\} $.
Without loss of generality, let each training sample in the dataset be an input-output pair $(\mathbf{x},\mathbf{y})$, where $\mathbf{x}$ is the feature and $\mathbf{y}$ is the label.
The set of parameters for the FL model is denoted as $\mathbf{w}$.
For each sample $i$, the loss function of the machine learning problem is defined as ${f_i}(\mathbf{w})=l({\mathbf{x}_i},{\mathbf{y}_i}|\mathbf{w})$.
Therefore, the loss function for device $j$ on the mini-batch $b_j$, a randomly sampled subset of $D_j$, can be written as ${f_{b_j}}(\mathbf{w})$.
The goal of device $j$ is to minimize the loss on each mini-batch:
\begin{equation}
	\label{eq:localProb}
	\min F_{j}(\mathbf{w})=\mathbb{E}_{b_{j}\sim {D_{j}}}  f_{b_{j}}(\mathbf{w}).
\end{equation}
By applying the gradient descent algorithm on the mini-batch, the local model of device $j$ can be updated according to:
\begin{equation}
	{{\bf{w}}_j} \leftarrow {{\bf{w}}_j} - {\mu}_j \nabla {f}_{b_j}( {{{\bf{w}}_j}} ),
\end{equation}
where ${\mu}_j$ is the learning rate of this device.
Then, $E$ epochs for local dataset $D_{j}$ are executed to train the local model.

In addition, the Federated Averaging algorithm \cite{mcmahan2017communicationefficientlearning} is adopted to aggregate the updated local models uploaded from the selected devices.
Then the loss function of shard \#1 on decentralized datasets can be expressed as:
\begin{equation}
	G_{s1}(\mathbf{w}) = \sum\limits_{j = 1}^m {\frac{{\left| {{D_j}} \right|}}{D}} {F_j}(\mathbf{w}),
\end{equation}
where $m (m \le n)$ is the number of valid models that pass the validation during the consensus, and $D = \sum\nolimits_{j = 1}^m {|{D_j}|}$ is the total size of the datasets used in this shard training round.
As the IoT devices selected in round $k$ upload their updated local models, the model parameters of shard \#1, ${\bf{w}}_{s1}$, are updated through the weighted aggregation of all updated local models' parameters, i.e.,
\begin{equation}
	\label{eq:shardAggregate}
	{\bf{w}}_{s1}(k) = \sum\limits_{j = 1}^m {\frac{{|{D_j}|{{\bf{w}}_j}(k)}}{D}}.
\end{equation}

\subsection{FL Process}
\label{subsec:procedure}

\begin{figure}[t]
	\centering
	\includegraphics[width=0.48\textwidth]{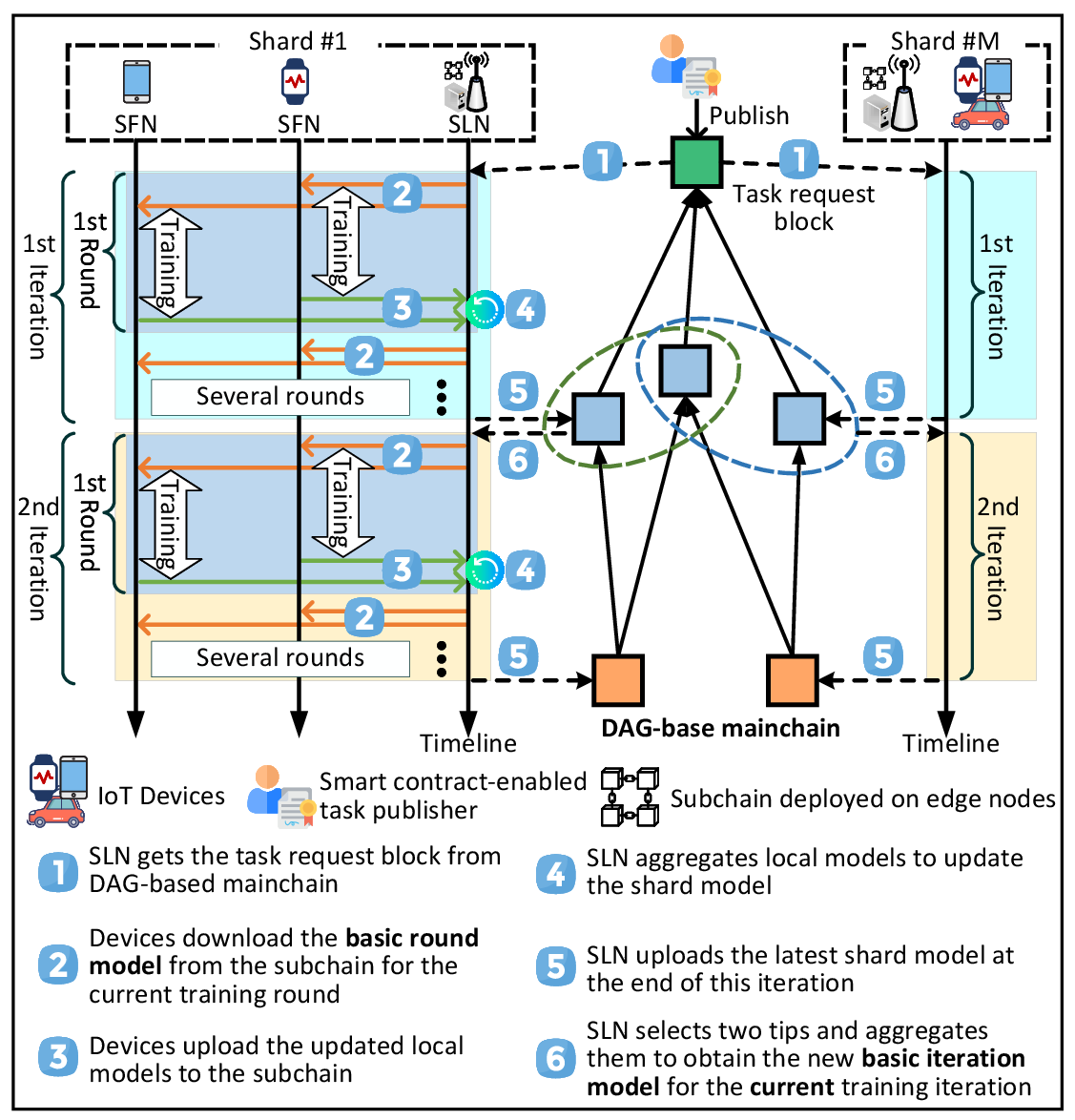}
	\caption{Overview of the FL process in ChainFL.}
	\label{fig:procedure}
\end{figure}

To complete the FL task in a decentralized manner, we define the operation process and design a set of interaction rules to orchestrate the IoT devices and edge nodes in ChainFL, as shown in Fig. \ref{fig:procedure}.
It is important to highlight that ChainFL incorporates two distinct types of transactions: \emph{subchain transactions} and \emph{mainchain transactions}.
The former is created by IoT devices and SLNs and spreads within a specific shard, while the latter is created by SLNs and spreads in the mainchain network.
Further details about this procedure are given as follows.

\textbf{Phase 1: Task Publication.}
To initiate an FL task, the task requester signs a smart contract that contains all the requirements of the task, such as the structure and parameters of the initial model, shard training configurations, and completion conditions.
The smart contract then generates the task request transaction (denoted as $g_0$) on the mainchain, which encapsulates the task requirements and a test dataset provided by the task requester.
Meanwhile, the smart contract triggers the corresponding shard network(s) to start the training task.

\textbf{Phase 2: Shard Training.}
SLNs in activated shard networks retrieve $g_0$, and the task information extracted from $g_0$ is encapsulated into a subchain transaction. 
This transaction is then recorded in the distributed ledger of each shard to initiate the training process.
The details of Phase 2 are described as follows:

\begin{algorithm}[t]
	\caption{\textbf{ShardTrainingIteration}. $\mathbf{w}_{bim}$: basic iteration model, $\mathbf{w}_{brm}$: basic round model, $\mathbf{w}_{s}$: shard model, $R$: the number of training round in each iteration.}
	\label{alg:oneShardTrainingIteration}
	\begin{algorithmic}[1]
		\SUB {\textbf{Each triggered SLN executes:}}
		\STATE obtain $\mathbf{w}_{bim}$ from mainchain for the current shard training iteration
		\STATE $\mathbf{w}_{brm} = \mathbf{w}_{bim}$, $\mathbf{w}_{s} = \mathbf{w}_{bim}$, $r = 0$
		\WHILE{$r < R$}
		\STATE encapsulate $\mathbf{w}_{brm}$ and publish to the subchain
		\STATE select and trigger devices
		\STATE receive valid local models \textsl{\textsf{ // waiting devices update}}
		\STATE $\mathbf{w}_{s} \leftarrow$ aggregate local models according to (\ref{eq:shardAggregate})
		\STATE $\mathbf{w}_{brm} = \mathbf{w}_{s}$, $r = r+1$
		\ENDWHILE
		\STATE return $\mathbf{w}_{s}$
	\end{algorithmic}
	\begin{algorithmic}[1]
		\SUB {\textbf{Nodes in subchain execute:}}
		\STATE receive the transactions from devices
		\FOR {all received transactions}
		\STATE $A_{new} \leftarrow$ validate the accuracy of the model stored in the transaction
		\IF {$A_{new} > A_{\tau}$ }
		\STATE forward the transaction to SLN
		\ELSE
		\STATE mark invalid and discard
		\ENDIF
		\ENDFOR
	\end{algorithmic}
\end{algorithm}

\begin{enumerate}[1)]

	\item \emph{Device Selection:}
	      In each training round of one shard, SLN selects candidates for shard training based on the status of IoT devices, such as their local data profile and power status, which are reported periodically.
		  Only devices ready for training with sufficient battery and stable network coverage are chosen.
		  These devices are then authorized to access the subchain, download the basic round model for local training, and subsequently upload their updated models. 
	      It is worth noting that device selection in each shard operates independently.

	\item \emph{Local Update:}
		  Utilizing the basic round model obtained from the subchain, each device performs the local training process by engaging with its raw data to address problem (\ref{eq:localProb}). 
		  Once the predefined goals set in the smart contract, such as a certain number of local training epochs or a target evaluation metric convergence value, are met, the updated local model is transmitted to the relevant subchain node (SLN or SFN).

	\item \emph{Model Aggregation:}
		  As outlined in Algorithm \ref{alg:oneShardTrainingIteration}, subchain nodes first receive and validate local models against the test dataset.
	      The validity of these models is determined using the preset threshold $A_{\tau}$, which is typically aligned with the evaluation metric of the basic round model relevant to the current training round, such as accuracy for target recognition tasks or perplexity for natural language processing.
		  Subsequently, the SLN aggregates the valid local models following \eqref{eq:shardAggregate} to update the \emph{shard model}, which is then disseminated to the subchain as the \emph{basic round model} for the ongoing shard training iteration.
	      Due to the synchronous manner of the training within each shard, the aggregation of the shard model is triggered when sufficient IoT devices upload their local models within a specified period of time, otherwise, the round is discarded.
		  This entire procedure, encompassing device selection and model aggregation, constitutes \emph{a round of shard training}.
	      If the iteration continues, the updated shard model is packaged as a new subchain transaction and published to the subchain, forming the basis for the next shard training round.

\end{enumerate}

\begin{algorithm}[t]
	\caption{SLN Interact with Mainchain: Basic Iteration Model Building and Shard Model Submitting}
	\label{alg:LeaderExecutes}
	\begin{algorithmic}[1]
		\WHILE{true}
		\IF {the current is 1st iteration}
		\STATE $\mathbf{w}_{bim} \leftarrow$ extract the initial parameters from $g_0$
		\STATE ApproveSet = ($g_0$)
		\ELSE
		\STATE $(\mathbf{w}^{'}_1,\mathbf{w}^{'}_2,...,\mathbf{w}^{'}_\eta)\leftarrow$ choose $\eta$ tips from the DAG
		\STATE $(A_1^{'},A_2^{'},...,A_{\eta}^{'}) \leftarrow $ validate the accuracy of the model in each chosen tip
		\STATE $\mathbf{w}_{bim} \leftarrow$  $(\sum\limits_{i=1}^{\lambda } {\frac{\mathbf{w}_i^{'}}{\lambda}})$, aggregate $\lambda (\lambda < \eta) $ tips with the highest accuracy to build a basic iteration model
		\STATE ApproveSet = these $\lambda$ tips
		\ENDIF
		\STATE $\mathbf{w}_{new} \leftarrow$ \textbf{ShardTrainingIteration}($\mathbf{w}_{bim}$)
		\STATE $g \leftarrow$ package $\mathbf{w}_{new}$ and the ID of all transactions in ApproveSet as a mainchain transaction
		\STATE submit the $g$ to the mainchain
		\IF{stop signal received}
		\STATE \textbf{break}
		\ENDIF
		\ENDWHILE
	\end{algorithmic}
\end{algorithm}

\textbf{Phase 3: Shard Model Submitting and Basic Iteration Model Aggregating.}
Upon the completion of each shard training iteration, the most recent aggregated shard model is encapsulated within a mainchain transaction. 
This transaction is subsequently submitted to the mainchain by the SLN. 
Meanwhile, the new \emph{basic iteration model} $\mathbf{w}_{bim}$ is aggregated from the mainchain. 
This aggregation initiates the subsequent shard training iteration, as long as the training task is still ongoing.
It is important to emphasize that each shard independently coordinates its training, validates other transactions, and submits its trained shard model to the mainchain. 
This process forms the basis of the asynchronous transaction processing in the mainchain.
The details of these processes are shown in Algorithm \ref{alg:LeaderExecutes}.

In the decentralized architecture of ChainFL, direct generation of a global model is not feasible.
Hence, a smart contract is employed to monitor the latest DAG and execute operations analogous to those in the basic iteration model aggregation, as outlined in Algorithm \ref{alg:LeaderExecutes}, for periodic global model aggregation.
The selection of transactions for constructing the global model is governed by task-specific parameters, which are explicitly defined within the smart contract. 
Upon reaching the predetermined termination condition, the smart contract broadcasts a stop signal to all activated SLNs.
Subsequently, these SLNs conclude their training process after completing the current iteration.
Moreover, the task requester is endowed with the ability to aggregate the global model from any location with mainchain access, thus facilitating decentralized control. 
In addition, IoT devices, once authorized, can engage in training on the edge shard blockchain. 
This participation not only augments model accuracy but also allows for the convenient acquisition of the most up-to-date intelligent services, reducing the need for centralized infrastructure.

\subsection{ChainFL Consensus}
\label{sec:consensus}

As described in Section \ref{sec:architecture}, the Raft consensus and the DAG consensus are adopted in each subchain and the mainchain, respectively.

\subsubsection{\textbf{Raft Consensus}}
\label{subsec:raftConsensus}
In a Raft consensus-based subchain network, edge nodes are classified as leaders or followers.
While detailed leader selection is beyond the scope of this paper (see \cite{ongaro2014searchunderstandable} for more), 
managing leader failures, such as offline incidents, is crucial. 
Raft, a Crash Fault-Tolerance (CFT) protocol, ensures subchain continuity during leader crashes, detected via a heartbeat mechanism \cite{ongaro2014searchunderstandable}.
Once a leader crash is detected, leader candidates initiate an election process.
Specifically, in Raft, the maximum number of failed nodes that can be tolerated, denoted as $a$, is determined by the condition $b = 2a+1$, where $b$ represents the total number of edge nodes present within a shard.
As shown in Algorithm \ref{alg:oneShardTrainingIteration}, followers have the responsibility of validating received transactions (updated local models) and forwarding the valid ones to the leader.
The leader then arranges these transactions in chronological order based on their generation time.
Once the cumulative size of the transactions reaches a threshold or the designated period ends, the leader creates a block and broadcasts it to all followers.
Followers approve the block after confirming the signatures and verifying the transactions it contains.
Then, consensus on this block is reached when the leader receives positive responses from at least half of all followers.

By partitioning a large-scale blockchain network into multiple independent shards, the system throughput is scaled effectively with the help of parallel consensus and separate data storage.
These approaches localize most data synchronization to individual shards rather than the entire network, significantly reducing communication rounds and expediting transaction processing. 
In addition, local models are stored exclusively in the ledger of their respective shards, substantially decreasing the data storage requirements for the blockchain nodes.
Furthermore, the influence of stragglers is limited to their own shards, which prevents network-wide impact.

\subsubsection{\textbf{DAG Consensus-based Virtual Pruning}}

\begin{figure}[t]
	\centering
	\includegraphics[width=0.48\textwidth]{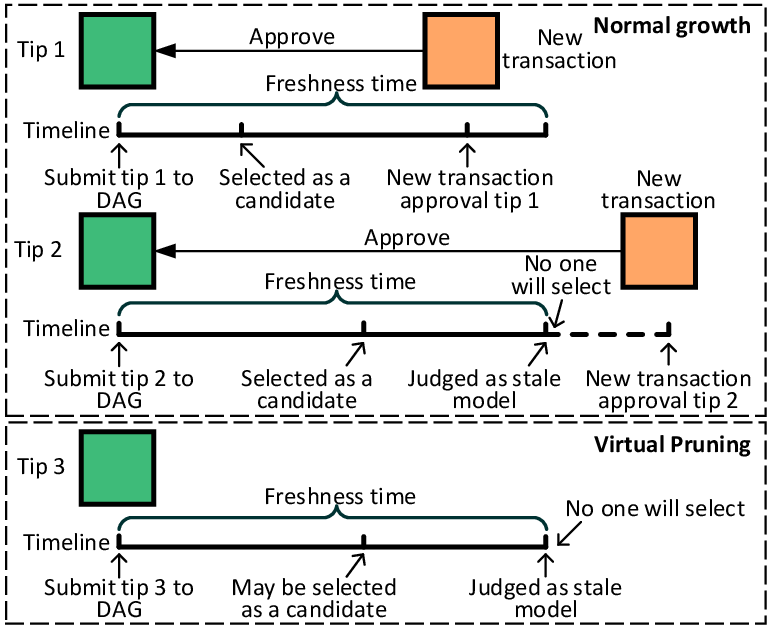}
	\caption{Three situations in the lifecycle of each tip.}
	\label{fig:virtualPruning}
\end{figure}

As outlined in Algorithm \ref{alg:LeaderExecutes}, SLNs utilize a set of $\lambda$ tips for the construction of the basic iteration model. 
Subsequently, these selected tips are approved by the updated shard model, which itself is trained from the basic iteration model in the current iteration, and is thereafter encapsulated in new mainchain transactions.
However, the mainchain might include two types of abnormal transactions: malicious transactions from malicious shards and stale transactions comprising outdated models from stragglers.
The tip selection, as presented in Algorithm \ref{alg:LeaderExecutes}, effectively detects such abnormal transactions by merging a voting system with mainchain transaction accuracy checks.
Transactions characterized by low accuracy, for instance, are more likely to be ignored, thus precluding their utilization in the aggregation of the basic iteration model. 
In Section \ref{sec:ProbabilityAnalysis}, we develop a probabilistic model to quantify the likelihood that the basic iterative model incorporates tips with low accuracy.
In DAG-based consensus, the approval rate for abnormal transactions is lower compared to normal ones, but these unapproved transactions are retained as tips in the graph of the mainchain. 
Over time, the growing share of abnormal transactions among tips increases the risk of their selection by SLNs for the basic iteration model.

To address this issue, we set a waiting period called \emph{freshness time} in the mainchain to eliminate the effect of abnormal transactions.
This freshness time is applied independently to each tip and begins counting once the tip is received by the mainchain node.
As depicted in Fig. \ref{fig:virtualPruning}, each tip undergoes one of three potential scenarios during its lifecycle.
For a tip to be approved by other transactions, it must be selected as a candidate by at least one SLN within its freshness time.
Tips failing to be chosen within their freshness time are subsequently disregarded by all SLNs.
This mechanism of tip selection, coupled with the enforcement of freshness time, effectively mitigates the influence of abnormal transactions, leading to a virtual pruning of the DAG.
In the mainchain, a transaction is deemed to have reached consensus when it secures approval from a sufficient number of other transactions, denoted as $N$, either directly or indirectly. 
Notably, while each Raft-based shard is limited to handling crash faults, this DAG consensus process efficiently mitigates the negative impact of malicious devices and shards on FL tasks.

\subsubsection{\textbf{Probability Analysis for Tip Selection}}
\label{sec:ProbabilityAnalysis}

As outlined in Algorithm 2, SLNs select $\eta$ tips from the DAG mainchain and subsequently identify the first $\lambda$ (where $\lambda < \eta$) tips with the highest accuracy to build a basic iteration model.
To model this probability, we introduce an auxiliary parameter denoted as $A^{\prime}$ (where $A^{\prime}\ge A_{\tau}$) and assume that there exist $a$ tips with an accuracy equal to or greater than $A^{\prime}$ in the DAG mainchain.
The total number of the available tips in the DAG mainchain is denoted as $I$.
Then, we can determine the probability of the $b$ (where $b \le I-a$) tips whose accuracy is below $A^{\prime}$ in selected $\eta$ tips, and the probability is
\begin{equation}
	P\left( b \right) =\frac{C_{I-a}^{b}C_{a}^{\eta -b}}{C_{I}^{\eta}}.
\end{equation}

In addition, the goal of tip selection during the interaction between the SLN and the DAG mainchain is to prioritize the tips with higher accuracy.
If the number of tips with an accuracy lower than $A^{\prime}$ in the selected $\eta$ tips is greater than $\eta-\lambda$, which is equivalent to $b>\eta-\lambda$, the basic iteration model will utilize tips with an accuracy below $A^{\prime}$.
Therefore, we can derive the probability that tips with an accuracy lower than $A^{\prime}$ are used to build the basic iteration model, and the probability is
\begin{equation}
	P\left( b>\eta -\lambda \right) =1-P\left( b\le \eta -\lambda \right) =1-\sum_{b=0}^{\eta -\lambda}{P\left( b \right)}.
\end{equation}
When holding other parameters constant, it can be observed that the aforementioned probability decreases as the value of $\lambda$ decreases.
This trend indicates that with a lower $\lambda$, transactions of lower accuracy are more likely to be excluded from consideration.

\begin{figure}[t]
	\setlength{\abovecaptionskip}{0pt}
	\centering
	\includegraphics[width=0.48\textwidth]{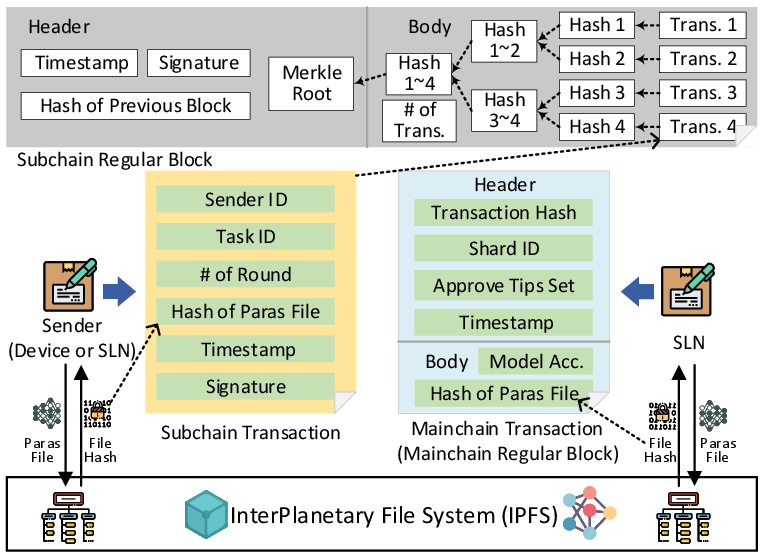}
	\caption{The format of transactions or/and blocks in ChainFL.}
	\label{fig:transINsubchain}
\end{figure}

\begin{figure*}[ht]
	\centering
	\subfigure[Implementation of ChainFL]{
		\label{fig:implementation}
		\includegraphics[width=0.4\textwidth]{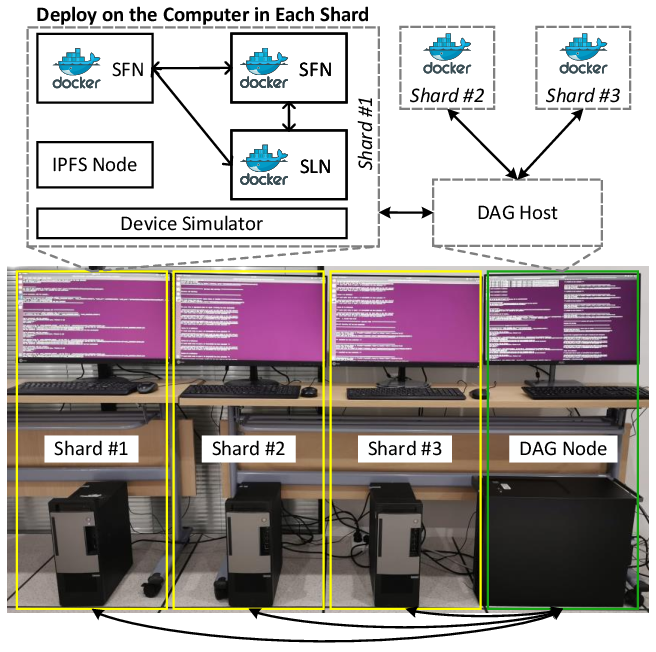}}
	\hspace{4mm}
	\subfigure[Function modules deployed on each entity.]{
		\label{fig:functions}
		\includegraphics[width=0.5\textwidth]{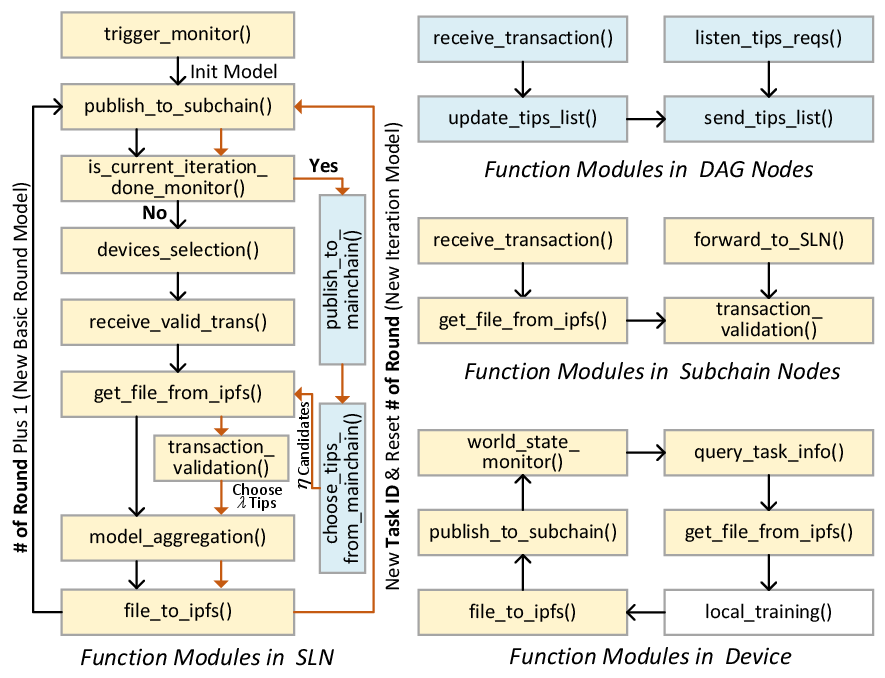}}
	\caption{The implementation of ChainFL in the real environment with the function modules deployed on each entity.}
	\label{fig:imp_funs}
\end{figure*}

\section{Implementation}
\label{sec:implementation}

In this section, we detail the practical deployment of ChainFL, which includes the off-chain storage scheme, Hyperledger Fabric-based subchain, and modified DAG-based mainchain.
The formats of transactions and blocks in ChainFL are presented in Fig. \ref{fig:transINsubchain}.
In addition, the real-world implementation of ChainFL, along with the function modules operational in each entity, are depicted in Fig. \ref{fig:implementation} and Fig. \ref{fig:functions}, respectively. 
The implementation of ChainFL is available on GitHub\footnote{https://github.com/shuoyuan/ChainsFL-implementation}.

\subsection{Off-Chain Storage Scheme}
Blockchain storage schemes typically fall into two categories: 1) full on-chain storage, where all data is directly stored in the blockchain ledger, and 2) off-chain storage, where data is stored in an external file system with only a unique identifier in the ledger ensuring immutability. 
Given the constraints of block size in Fabric, storing large data streams within the main body of the block is impractical. Consequently, our implementation adopts an off-chain storage approach.
For managing off-chain data, we utilize the InterPlanetary File System (IPFS) \cite{benet2014ipfscontent}, a private peer-to-peer file system. 
Upon adding a file to IPFS, it generates a unique hash value representing the file content. 
This hash value not only facilitates the reconstruction of the Merkle tree of file pieces of the parameter file but also enables the retrieval of the entire file \cite{benet2014ipfscontent}. 
In our system, the blockchain ledger stores only this hash value, not the parameter file itself.
To efficiently handle the parameter files within the training process, all IoT devices and blockchain nodes in ChainFL are integrated into the IPFS network. 
Interaction with IPFS is managed through two function modules deployed on each entity: \textsf{file\_to\_ipfs()} and \textsf{get\_file\_from\_ipfs()}, as illustrated in Fig. \ref{fig:functions}.

\subsection{Hyperledger Fabric-based Subchain}

To facilitate the implementation of subchains, we establish a Raft blockchain environment utilizing Hyperledger Fabric \cite{hyperledgerfabric} (referred to as Fabric), which incorporates the Raft ordering consensus mechanism. 
Fabric, a permissioned distributed ledger technology platform, is well-suited for the consortium-based structure of subchains in ChainFL. 
The Public Key Infrastructure (PKI)-based membership management in Fabric provides robust control over IoT device access, thus enabling efficient device selection.
In addition, a smart contract (called \emph{chaincode}) refined from \emph{sacc} \cite{hyperledgerfabric} is employed for processing transactions within subchains.
The format for recording local or shard model information in the subchain ledger is illustrated in Fig. \ref{fig:transINsubchain}.
Each subchain transaction includes a \textsf{Sender ID}, which is a unique identifier determined by the identity of the transaction issuer, such as an IoT device ID or an SLN ID.
The \textsf{Task ID}, assigned by the SLN at the initiation of a \emph{shard training iteration}, and the \textsf{\# of Round}, indicating the index of the current training round within an iteration, are also integral components.
Finally, the \textsf{Hash of Paras File} represents the hash value of a model file in IPFS, serving as a uniform resource identifier for file location within IPFS.

To augment the functionality of Fabric nodes in ChainFL, specialized function modules like \textsf{transaction\_validation()} and \textsf{model\_aggregation()} have been developed and integrated.
These modules, not inherent to the original Fabric, facilitate model validation and the aggregation of models and tips, as depicted in Fig. \ref{fig:functions}. 
In addition, functions pertinent to mainchain interactions, such as \textsf{publish\_to\_mainchain()} and \textsf{choose\_tips\_from\_mainchain()}, are also incorporated within Fabric nodes.
The scheduling rules and order among these modules are meticulously detailed in Fig. \ref{fig:functions}.
As described in Section \ref{subsec:procedure}, a new basic iteration model is aggregated from the mainchain upon the completion of the current iteration.
The progression of iterations is tracked via \textsf{\# of Round}, with iteration completion denoted by reaching the predefined round number in the task requirements. 
Moreover, each distinct shard training iteration is associated with a unique \textsf{Task ID}. 
Following the construction of a new basic iteration model, the SLN is tasked with generating a new \textsf{Task ID} and resetting \textsf{\# of Round} for the subsequent shard training iteration.

Due to the limited amount of hardware, a complete Fabric-based subchain containing one leader and two followers is configured in a single PC\footnote{The Raft ordering consensus of Fabric necessitates the deployment of at least three nodes (one leader and two followers) \cite{hyperledgerfabric}.}.
In addition, the training process of IoT devices served by this subchain is also simulated and executed on the same PC.
Consequently, this configuration essentially constitutes a single shard network within ChainFL, comprising the PC and its associated IoT devices.

\subsection{Modified DAG-based Mainchain}

We developed a modified DAG-based mainchain in Python to facilitate information exchange with shards.
The function modules specific to the mainchain nodes are depicted in Fig. \ref{fig:functions}.
The mainchain node maintains an updated tip list, shared with SLNs upon request.
The communication between SLNs and the mainchain node, predominantly for request-response interactions, is implemented through socket communication protocols.
Moreover, the tip list undergoes updates in two scenarios: first, when an SLN contributes a new transaction to the mainchain, and second, upon the detection of an abnormal tip through the virtual pruning mechanism.
The format of transactions in the mainchain is depicted in Fig. \ref{fig:transINsubchain}.
To accelerate and streamline the execution of massive experiments, the mainchain is deployed on a single computer (one node) in our real experimental setup, rather than on multiple distributed nodes\footnote{The deployment scheme does not impact interactions between SLNs and the DAG, ensuring that the performance of federated learning via ChainFL remains undisturbed.}.

\section{Experimental Evaluations} %
\label{sec:evaluation}

In this section, we evaluate the performance of ChainFL in terms of convergence and robustness against model attacks of malicious devices or shards.

\subsection{Baselines and Settings}

To evaluate the performance of ChainFL, we run two tasks: realistic object classification using Convolutional Neural Networks (CNNs) as Task 1, and neural language processing with Gated Recurrent Units (GRUs) as Task 2. 
Task 1 utilizes the MNIST image dataset \cite{lecun1998gradientbasedlearning}, while Task 2 employs the English language dataset Penn Treebank \cite{marcus1993buildinglarge}. 
For Task 1, we adopt a non-IID data setting where the MNIST training set is divided into 100 groups sorted by digit labels, with each device receiving one group. 
In Task 2, the Penn Treebank dataset is shuffled and randomly split into 100 subsets without replacement, and then each device is allocated one subset.

\begin{table}[t]
	\centering
	\scriptsize
	\caption{Common Experimental Settings.}
	\label{tab:expSettings}
	\begin{tabular}{llll}
		\hline
		Parameter         & Symbol       & Task 1               & Task 2               \\
		\hline
		Dataset           & $D$          & MNIST                & Penn Treebank        \\
		Dataset size      & $|D|$        & 70000                & 1036580              \\
		Model             & $\mathbf{w}$ & CNN                  & GRU                  \\
		\# of devices     & $n$          & 100                  & 100                  \\
		Learning rate     & $\mu$        & 1e-2                 & 1e-2                 \\
		\# of cand. tips  & $\eta$       & 3                    & 3                    \\
		\# of appr. tips  & $\lambda$    & 2                    & 2                    \\
		\# of shards      & $M$          & 3                    & 3                    \\
		Loss function     & $l$          & Cross Entropy Loss   & NLL Loss             \\
		Eval. metric      & $e_m$        & \makecell[c]{Acc=                           \\ $\frac{1}{n} \sum_{i=1}^{n} \phi\left({\mathbf{y}_i}, \hat{\mathbf{y}_i}\right)$}    & \makecell[c]{$PPL(x)=$ \\ $2^{-\sum_{x} p(x) \log \frac{1}{p(x)}}$} \\
		\# of dev./shard  & $S_{d}$      & $\{10,20,30\}$       & $\{10,20,30\}$       \\
		Mini-batch size   & $B$          & $\{10,20,30,40,50\}$ & $\{10,20,30,40,50\}$ \\
		Local epochs      & $E$          & $\{1,5,10,15,20\}$   & $\{1,5,10,15,20\}$   \\
		Malicious ratio   & $M_d$        & $\{0,0.1,0.2,0.3\}$  & $\{0,0.1,0.2,0.3\}$  \\
		\# of rounds/ite. & $R$          & $\{1,2,3\}$          & $\{1,2,3\}$          \\
		\hline
	\end{tabular}
\end{table}

In Task 1, we use the classic network of LeNet-5 which consists of two convolutional layers with max pooling and three fully connected layers. 
Task 2 simulates mobile keyboard scenarios in decentralized applications, where each text sample is embedded into a 300-dimensional vector for the GRU-based model, followed by a fully connected layer for next-word prediction, similar to \cite{ji2019learningprivate}. 
Table \ref{tab:expSettings} details common experimental settings, including different evaluation metrics ($e_m$) for different tasks.
In Task 1, for accuracy assessment, the function $\phi(\cdot)$ outputs 1 for correct model predictions ($\hat{\mathbf{y}_i}$) and 0 otherwise. 
Task 2 employs perplexity \cite{mikolov2011extensionsrecurrent}, a common metric for language models, where lower testing perplexity indicates higher accuracy and better performance, as detailed in Table \ref{tab:expSettings}.

Our extensive experiments compare various dimensions, including distributed training methods, mini-batch size scales, and local epoch numbers. 
We establish two baselines for comparison, with specific settings for baselines and ChainFL as follows:

\emph{FedAvg \cite{mcmahan2017communicationefficientlearning}:}
FedAvg, a synchronous federated optimization method, samples a fraction of devices in each iteration, with each device performing multiple local epochs to update its model. 
For a more fair comparison, the number of devices sampled per iteration equals the number of devices per shard ($S_d$).

\emph{AsynFL \cite{xie2019asynchronousfederated}:}
AsynFL is an asynchronous federated optimization method that updates the global model timely as the central server receives the updated local model from the device.
For detail, the global model $\mathbf{w}^{'}_{gm}$ is updated using the rule $\mathbf{w}^{'}_{new} \leftarrow \frac{1}{2}\mathbf{w}^{'}_{gm} + \frac{1}{2}\mathbf{w}^{'}_{lm}$ in each global epoch by using the updated local model $\mathbf{w}^{'}_{lm}$.

\emph{ChainFL:}
The training process of ChainFL in each shard takes a similar setting to FedAvg and performs in a decentralized way.
The basic iteration model of each shard is built asynchronously from the DAG according to Algorithm \ref{alg:LeaderExecutes}.
With three shards in ChainFL, $S_d$ non-overlapping devices are selected from 100 devices for each shard.

\begin{table*}[t]
	\centering
	\caption{Best Accuracy of Task 1 Under Different Experimental Settings of Mini-Batch Size (B) and \# of Local Epochs (E).}
	\label{tab:task1Sensitivity}
	\begin{tabular}{clllllllllll}
		\hline
		                      &                        & \multicolumn{10}{c}{Best Accuracy}                                                                                                                                                                              \\
		\multicolumn{1}{l}{}  &                        & \multicolumn{5}{c}{Stop@ \# of Global Epochs=150} & \multicolumn{5}{c}{Stop@ \# of Gradients=7000}                                                                                                              \\\hline
		                      & Mini-Batch Size (B)    & 10                                                & 20                                             & 30              & 40              & 50              & 10      & 20     & 30              & 40     & 50     \\ \hline
		\multirow{3}{*}{E=5}  & FedAvg                 & 0.9702                                            & 0.9603                                         & 0.9575          & 0.9552          & 0.9526          & 0.9663  & 0.9602 & 0.9552          & 0.954  & 0.9507 \\
		                      & AsynFL                 & 0.9021                                            & 0.8715                                         & 0.8511          & 0.8486          & 0.8147          & 0.9759  & 0.9756 & 0.9749          & 0.9724 & 0.9726 \\
		                      & ChainFL                & \textbf{0.9758}                                   & \textbf{0.9678}                                & \textbf{0.9683} & \textbf{0.9597} & 0.9507          & 0.9756  & 0.9678 & 0.9680          & 0.9545 & 0.9478 \\ \hline
		\multicolumn{1}{l}{}  & \# of Local Epochs (E) & 1                                                 & 5                                              & 10              & 15              & 20              & 5       & 10     & 15              & 20     &        \\ \hline
		\multirow{3}{*}{B=10} & FedAvg                 & 0.9632                                            & 0.9746                                         & 0.9704          & 0.9715          & 0.9715          & 0.9729  & 0.9619 & 0.9482          & 0.9383 &        \\
		                      & AsynFL                 & 0.8483                                            & 0.9021                                         & 0.8774          & 0.8898          & 0.8978          & 0.9759  & 0.9665 & 0.959           & 0.9508 &        \\
		                      & ChainFL                & \textbf{0.9701}                                   & \textbf{0.9758}                                & \textbf{0.9785} & \textbf{0.9780} & \textbf{0.9799} & 0.97565 & 0.9625 & \textbf{0.9389} & 0.8991 &        \\ \hline
	\end{tabular}
\end{table*}

\begin{figure}[t]
	\centering
	\subfigure[]{
		\label{fig:task1Bge}
		\includegraphics[width=0.22\textwidth]{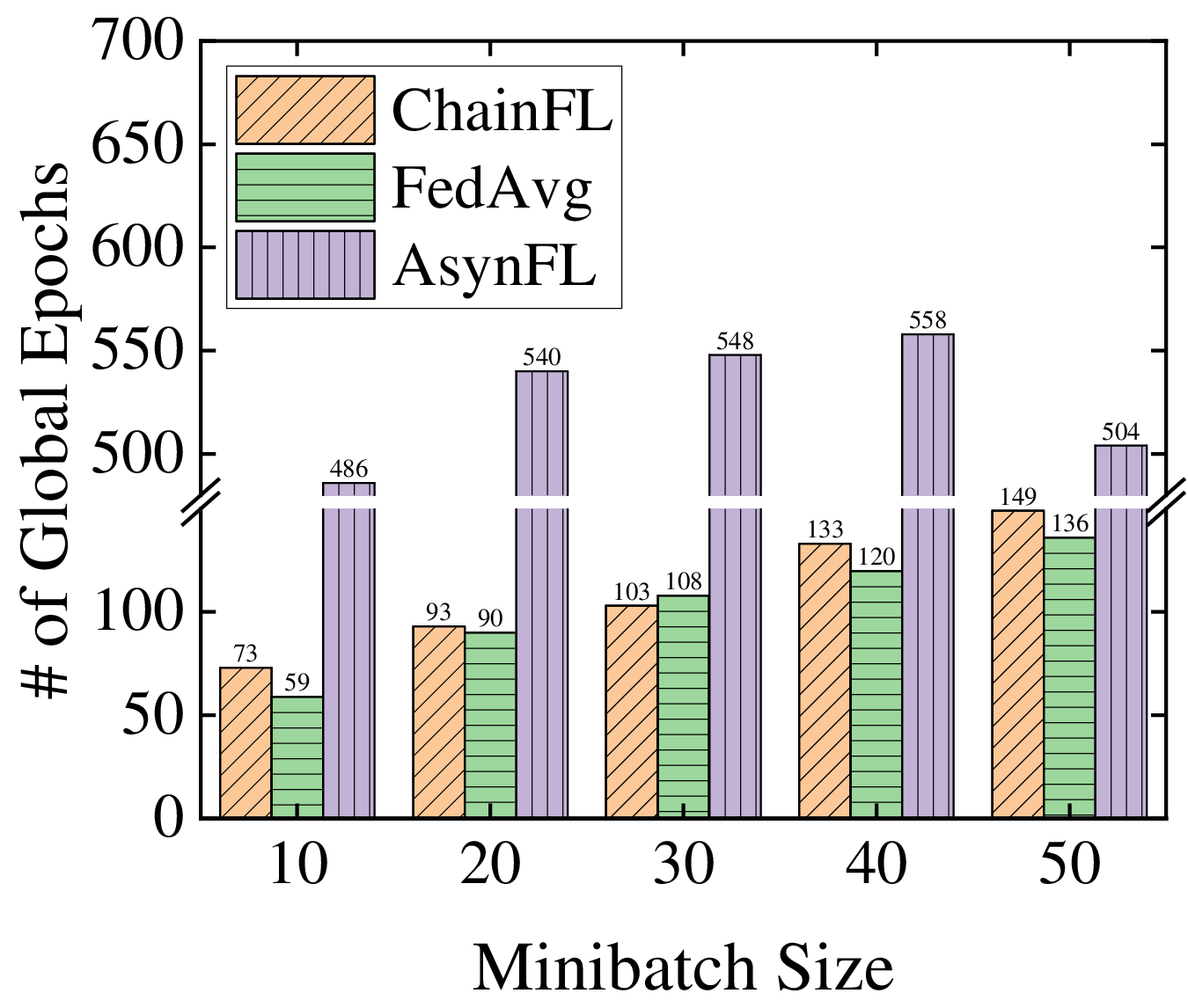}}
	\subfigure[]{
		\label{fig:task1Bgr}
		\includegraphics[width=0.21\textwidth]{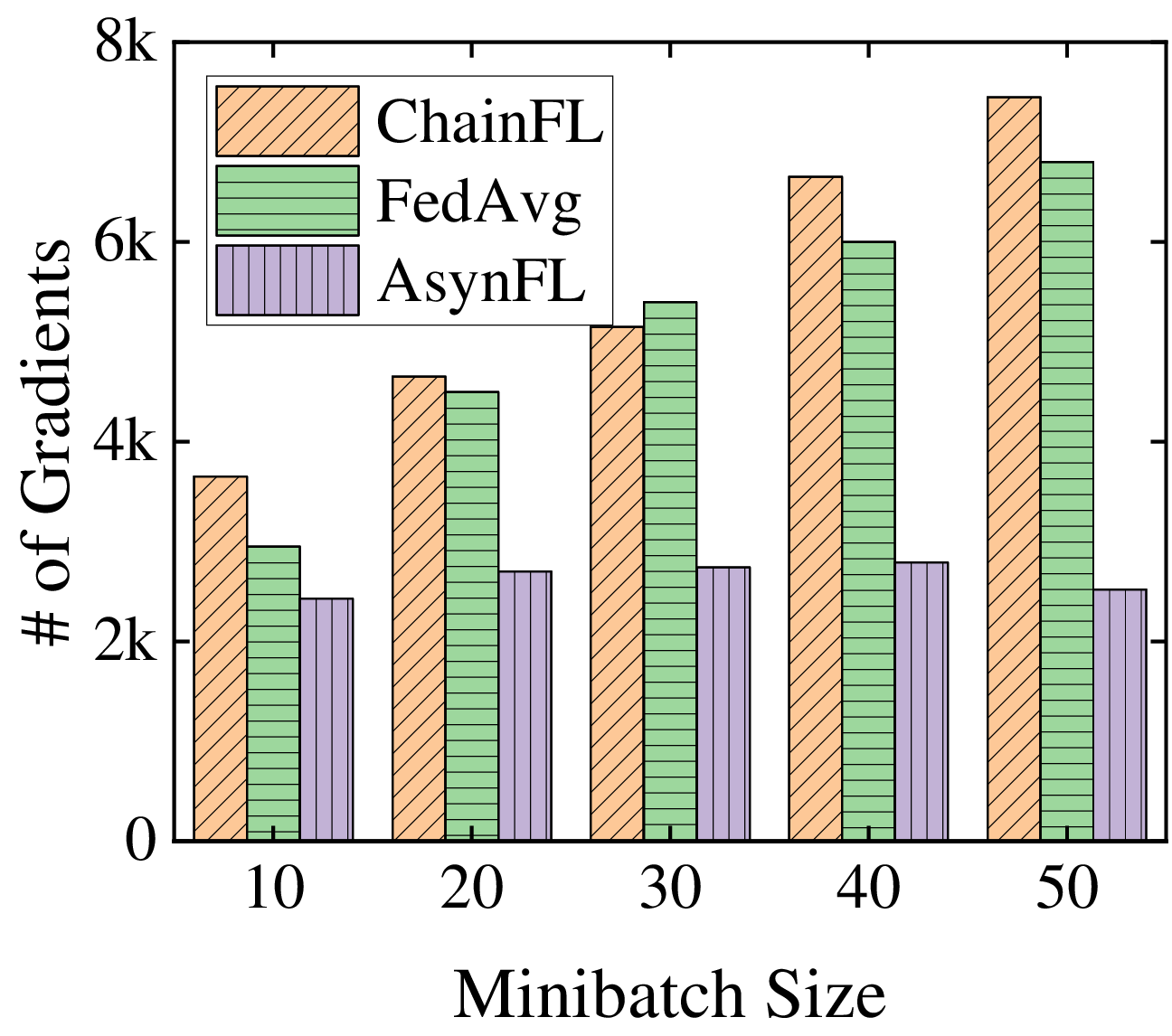}}

	\vfill

	\subfigure[]{
		\label{fig:task1Ege}
		\includegraphics[width=0.21\textwidth]{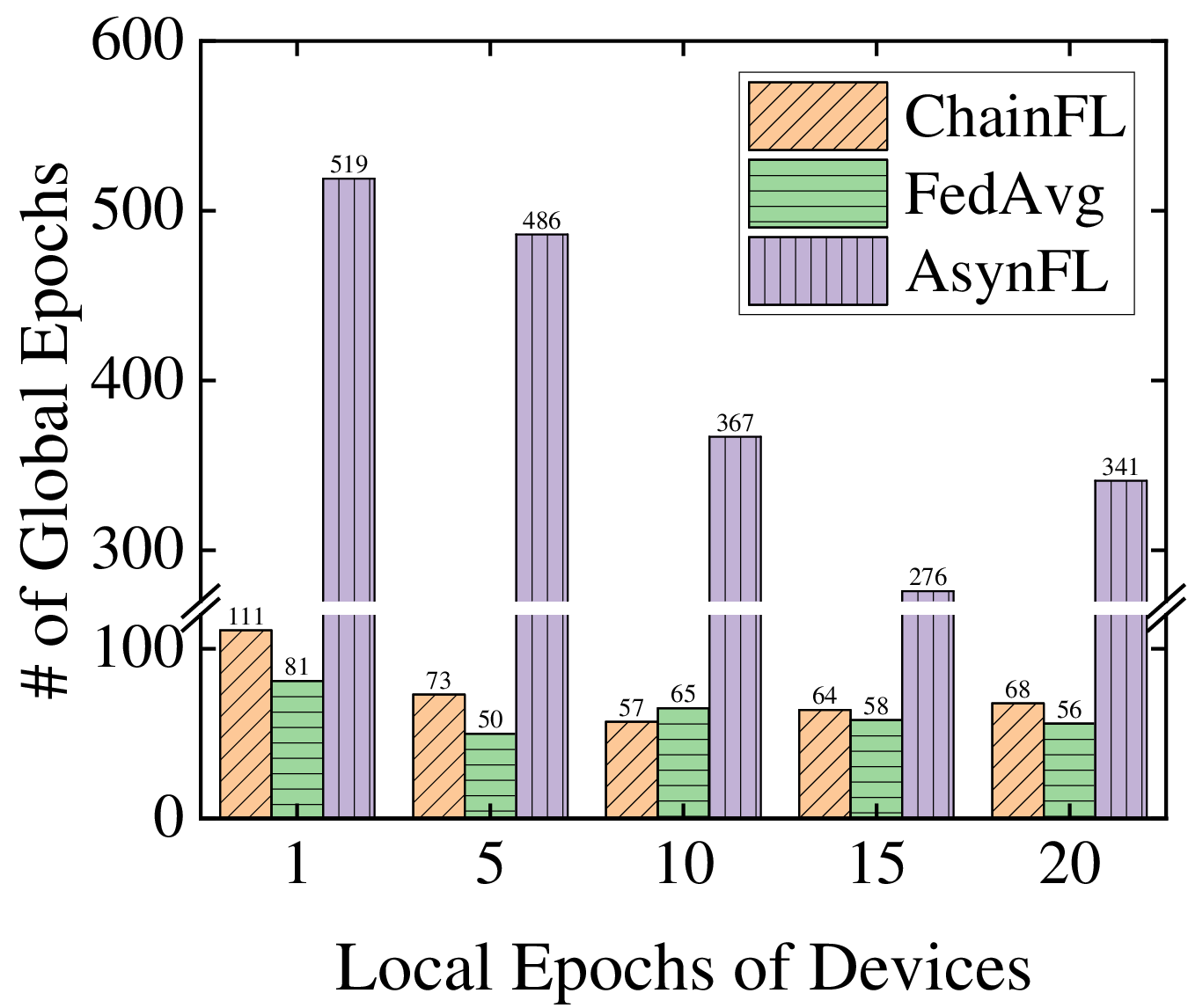}}
	\subfigure[]{
		\label{fig:task1Egr}
		\includegraphics[width=0.22\textwidth]{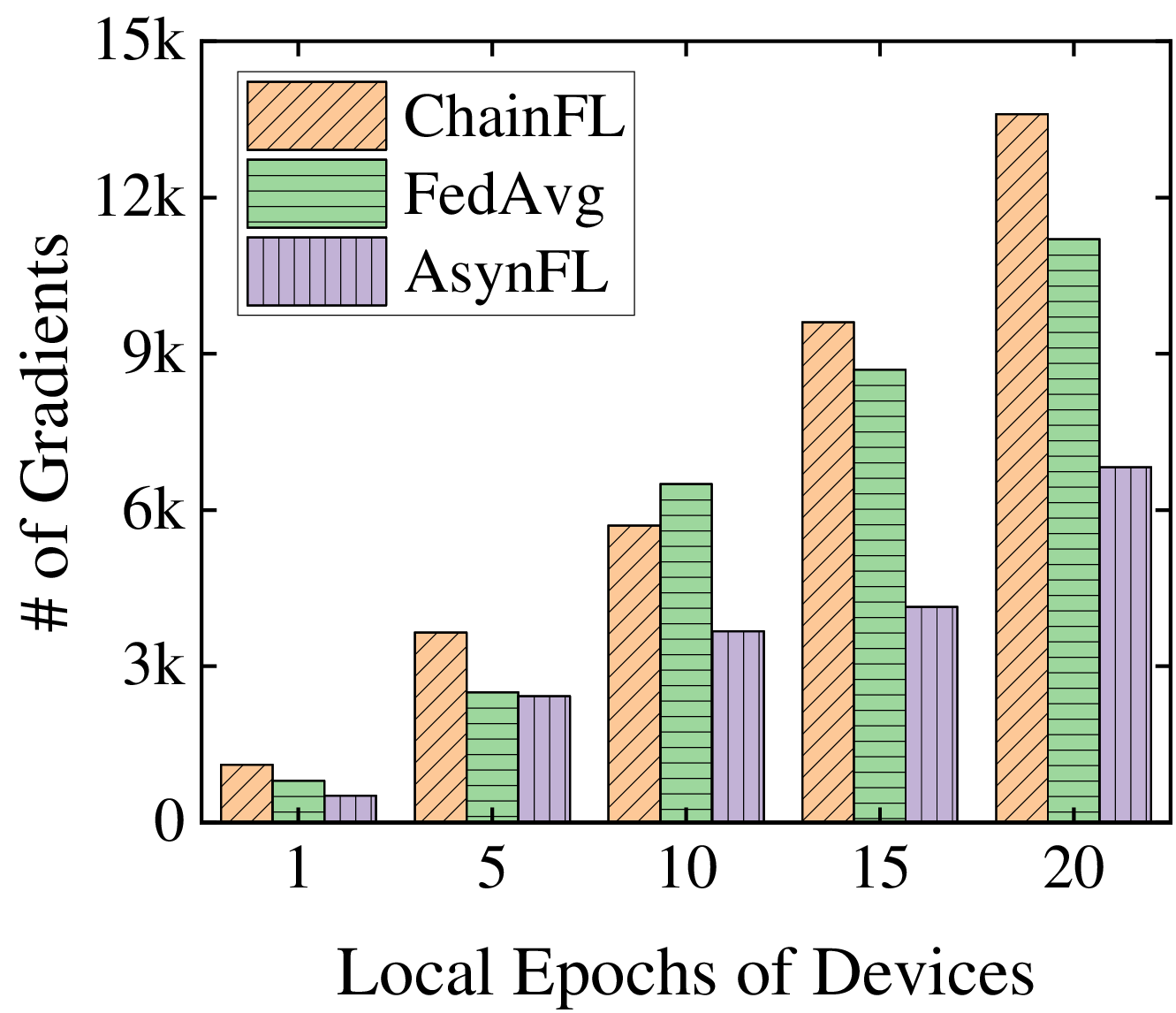}}
	\caption{Effect of the mini-batch size and local epochs of devices on \# of global epochs and \# of gradients with a preset threshold of the testing accuracy of 0.95 ($S_d=10$, $R=1$, $M_d=0$). (a) Global Epochs \emph{vs.} Mini-Batch Size. (b) Gradients \emph{vs.} Mini-Batch Size. (c) Global Epochs \emph{vs.} Local Epochs. (d) Gradients \emph{vs.} Local Epochs.}
	\label{fig:task1EwithGeANDGr}
\end{figure}

\begin{figure}[t]
	\centering
	\subfigure[]{
		\label{fig:task1AccGe}
		\includegraphics[width=0.22\textwidth]{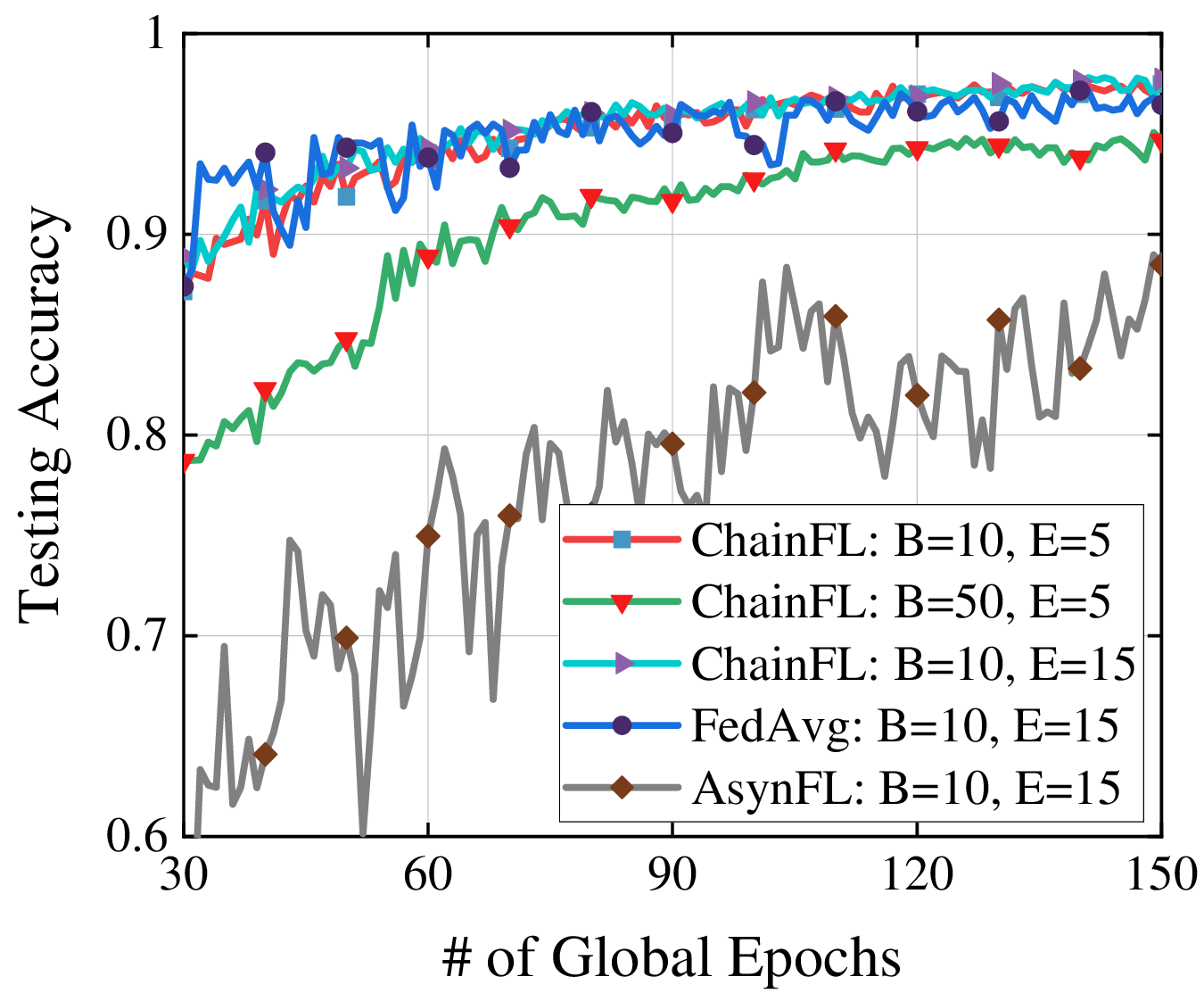}}
	\subfigure[]{
		\label{fig:task1LossGe}
		\includegraphics[width=0.21\textwidth]{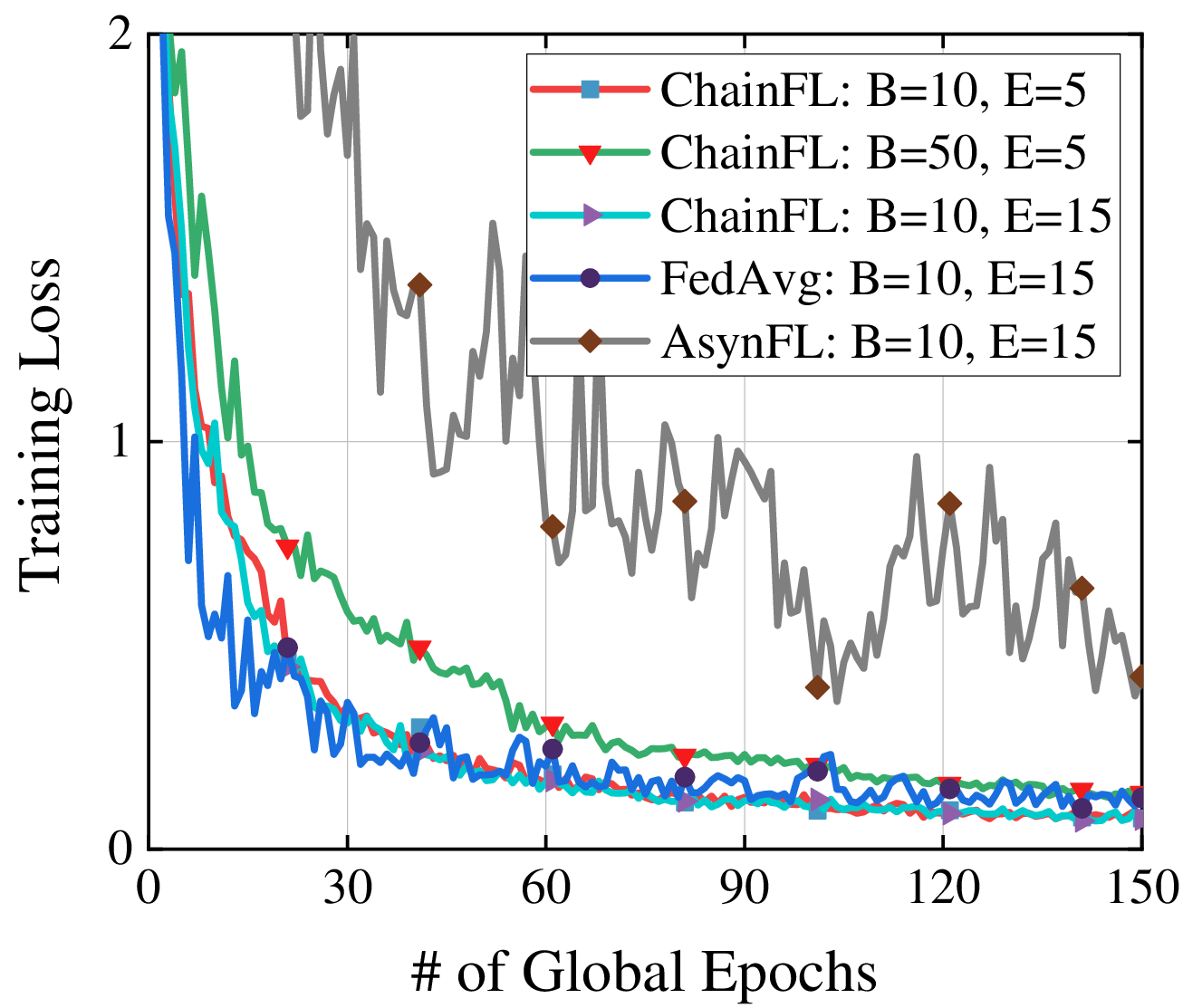}}

	\vfill

	\subfigure[]{
		\label{fig:task1AccGr}
		\includegraphics[width=0.22\textwidth]{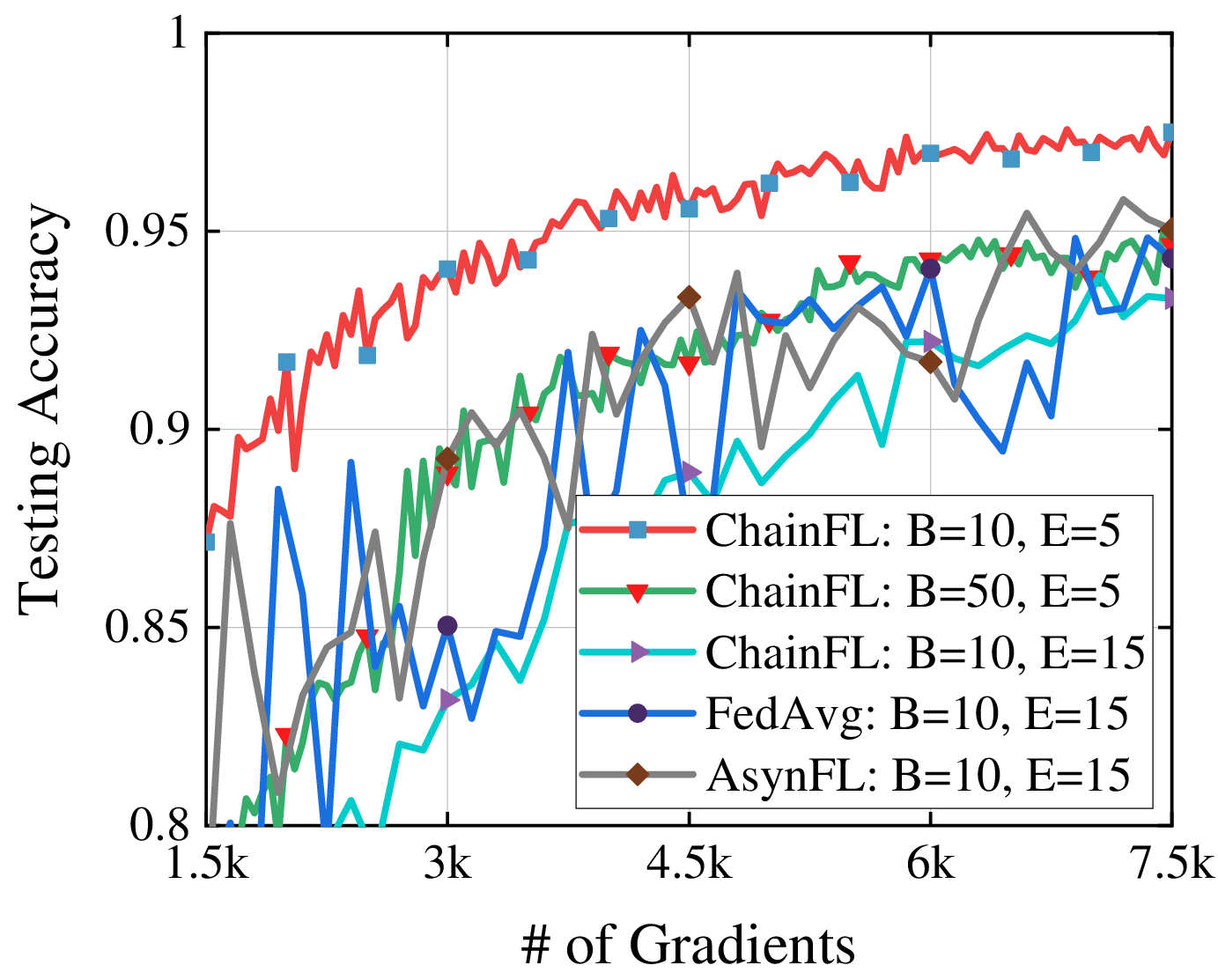}}
	\subfigure[]{
		\label{fig:task1LossGr}
		\includegraphics[width=0.21\textwidth]{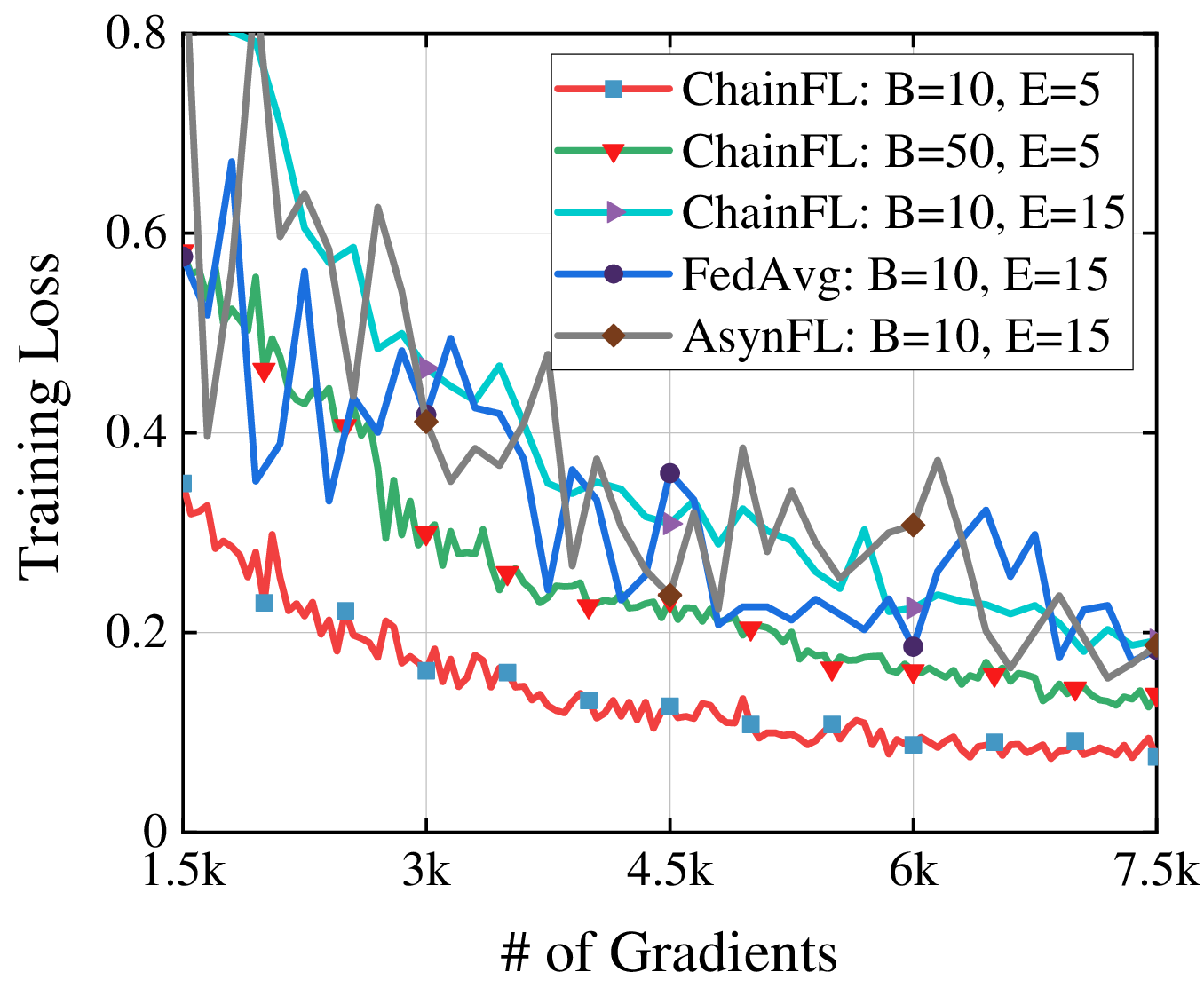}}
	\caption{Testing accuracy and training loss of Task 1 on two scales ($S_d=10$, $R=1$, $M_d=0$). (a) Accuracy \emph{vs.} Global Epochs. (b) Loss \emph{vs.} Global Epochs. (c) Accuracy \emph{vs.} Gradients. (d) Loss \emph{vs.} Gradients.}
	\label{fig:task1AccWITHLoss}
\end{figure}

\begin{figure}[t]
	\centering
	\includegraphics[width=0.48\textwidth]{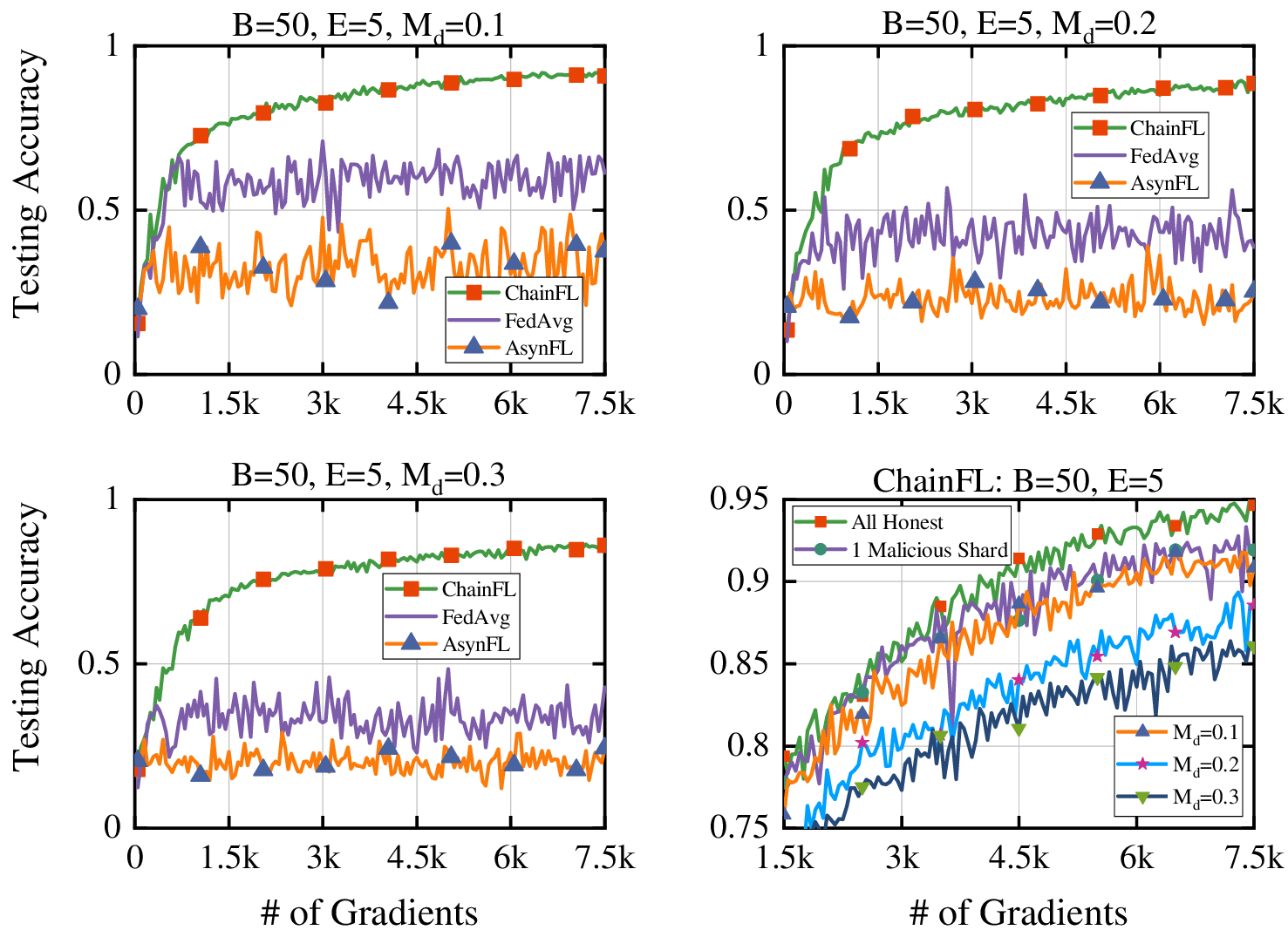}
	\caption{Effect of different malicious device ratio on the testing accuracy ($S_d=10$, $R=1$).}
	\label{fig:task1MN}
\end{figure}

It is evident that comparing the performance of AsynFL based on the number of global epochs may be unfair, given the varying number of devices selected in each training round.
To address this, we use two comparative approaches: metrics against the number of global epochs, and metrics against the number of gradients.
We treat each gradient trained in a local epoch of one device as a computational unit which allows performance evaluation at equivalent computational costs.
For example, using 50 gradients equates to selecting 10 devices per global epoch, with each device conducting 5 local epochs.
In ChainFL, the latest global model in the current mainchain is approximated by the basic iteration model aggregated by SLNs from tips of the mainchain due to similar model aggregation processes and the independent operation of each shard.
For all three paradigms, a local epoch of the device involves processing the entire local dataset.
It is also important to note that, due to the decentralized setup of ChainFL, communication rounds are not a suitable metric for fair comparison and are thus not evaluated in this study.

\begin{figure}[t]
	\centering
	\includegraphics[width=0.48\textwidth]{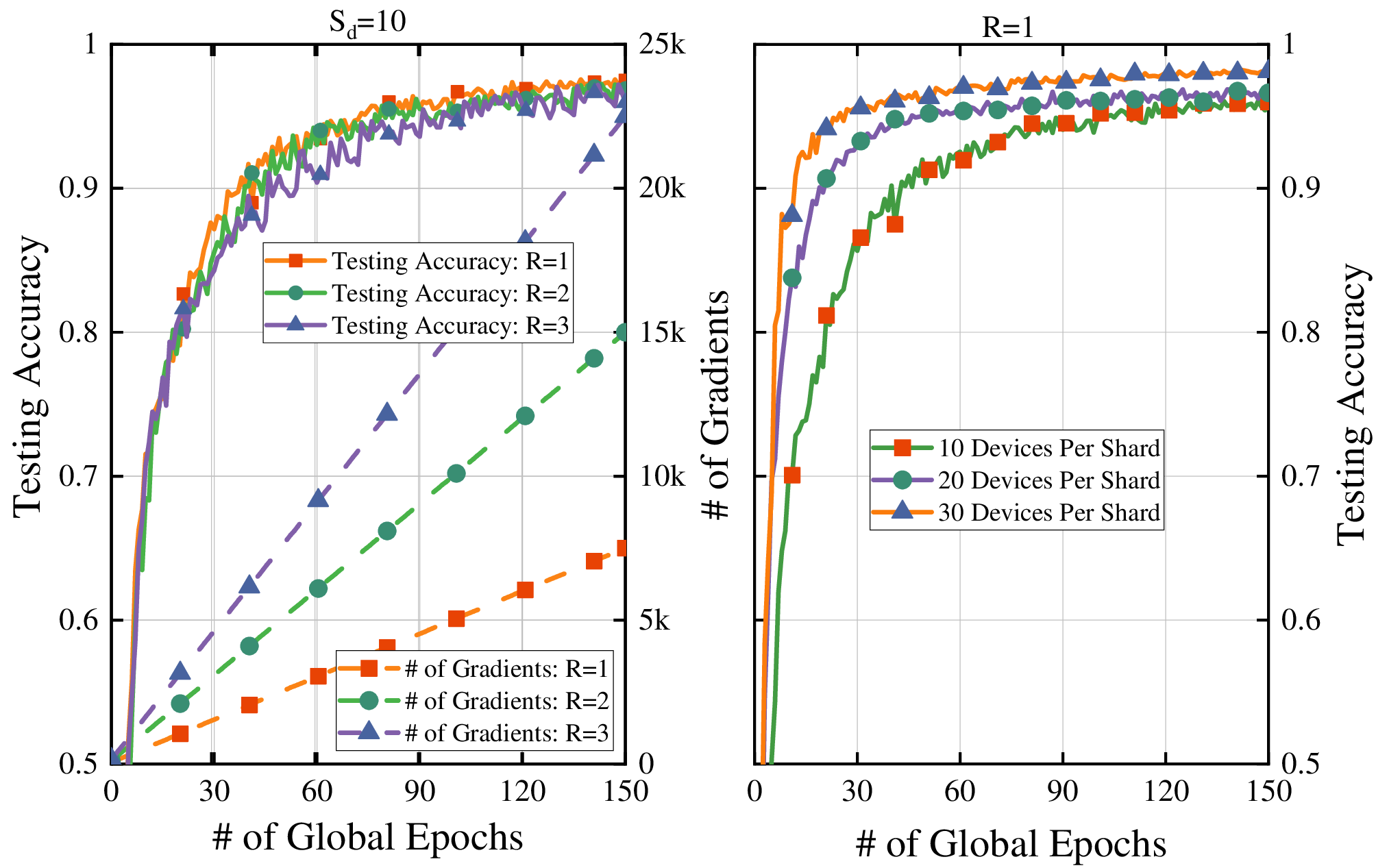}
	\caption{Effect of rounds per iteration and devices per shard on the testing accuracy ($B=10$, $E=5$, $M_d=0$).}
	\label{fig:task1roundCompare}
\end{figure}

\subsection{Experimental Results}

\begin{figure}[t]
	\centering
	\subfigure[]{
		\label{fig:task2Bge}
		\includegraphics[width=0.23\textwidth]{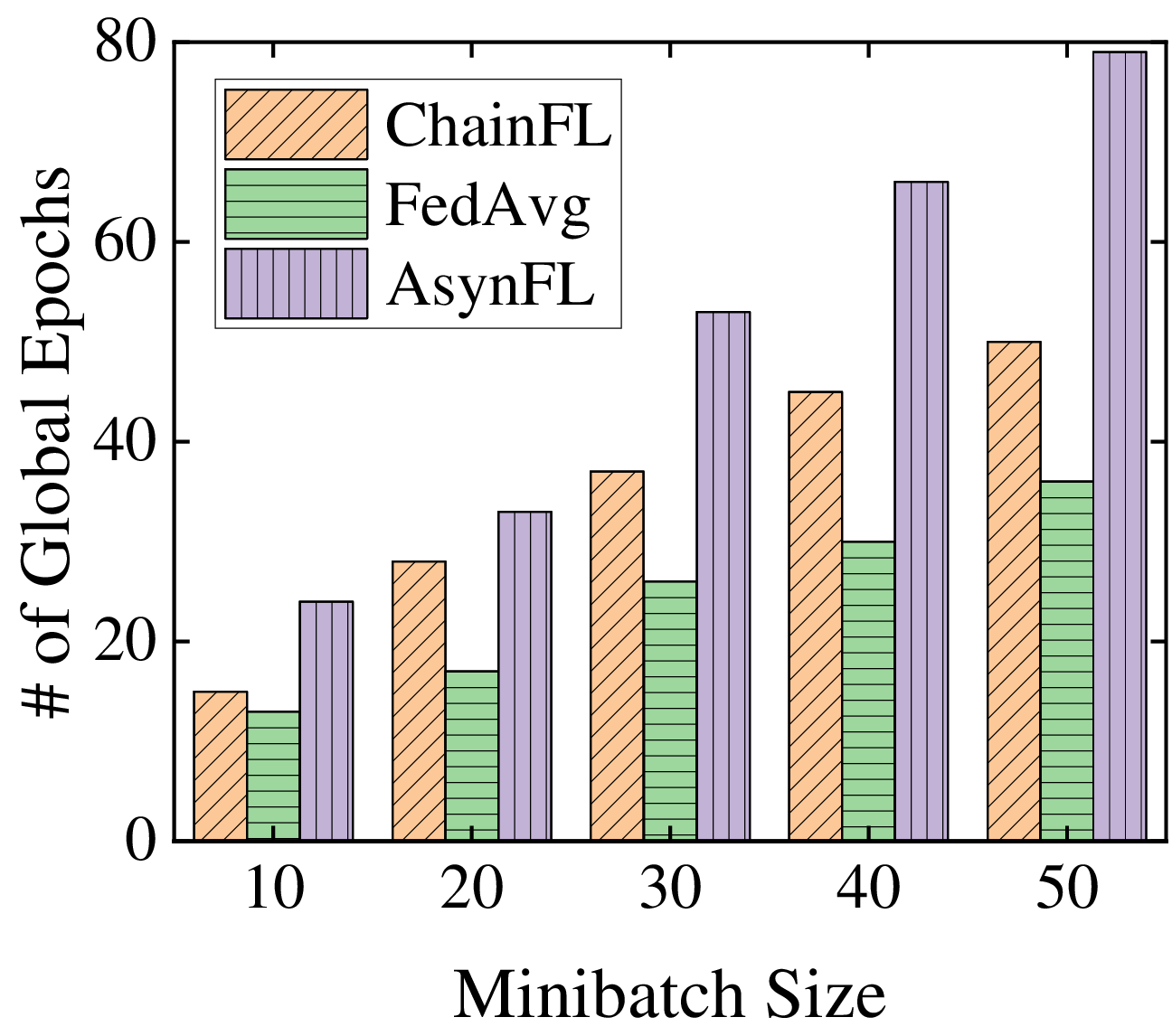}}
	\subfigure[]{
		\label{fig:task2Bgr}
		\includegraphics[width=0.23\textwidth]{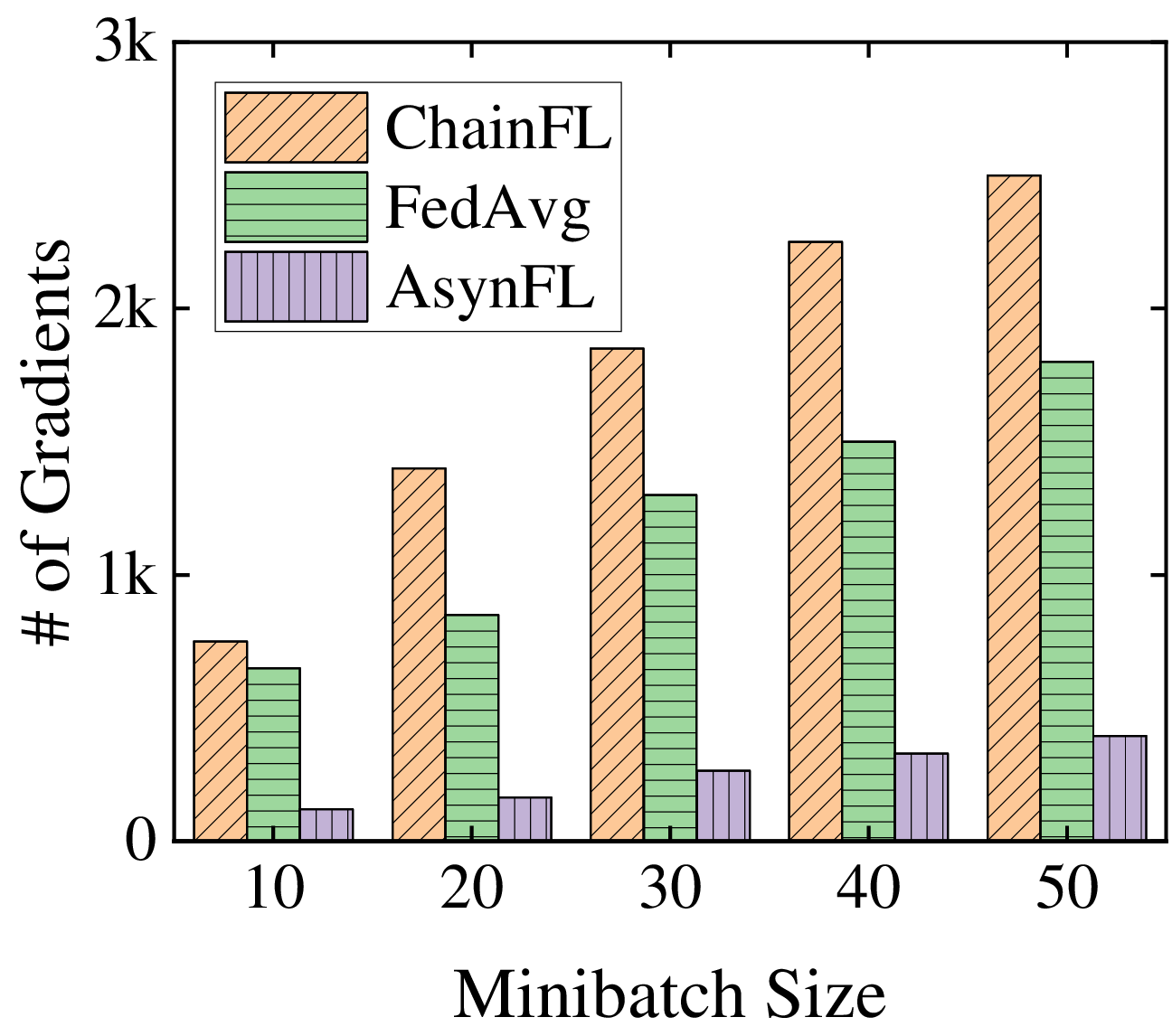}}

	\vfill
	\hspace{-0.2in}

	\subfigure[]{
		\label{fig:task2Ege}
		\includegraphics[width=0.23\textwidth]{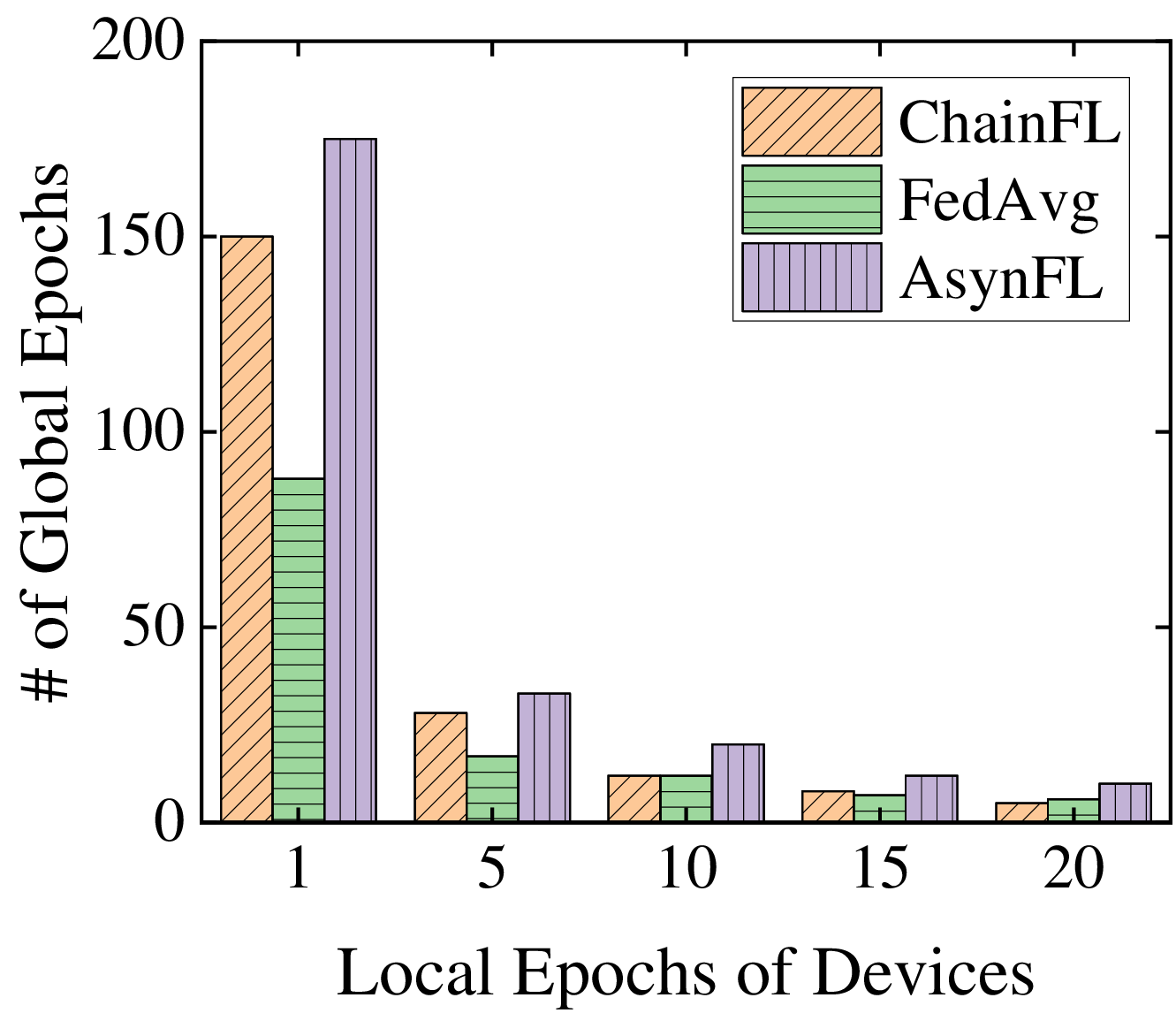}}
	\subfigure[]{
		\label{fig:task2Egr}
		\includegraphics[width=0.23\textwidth]{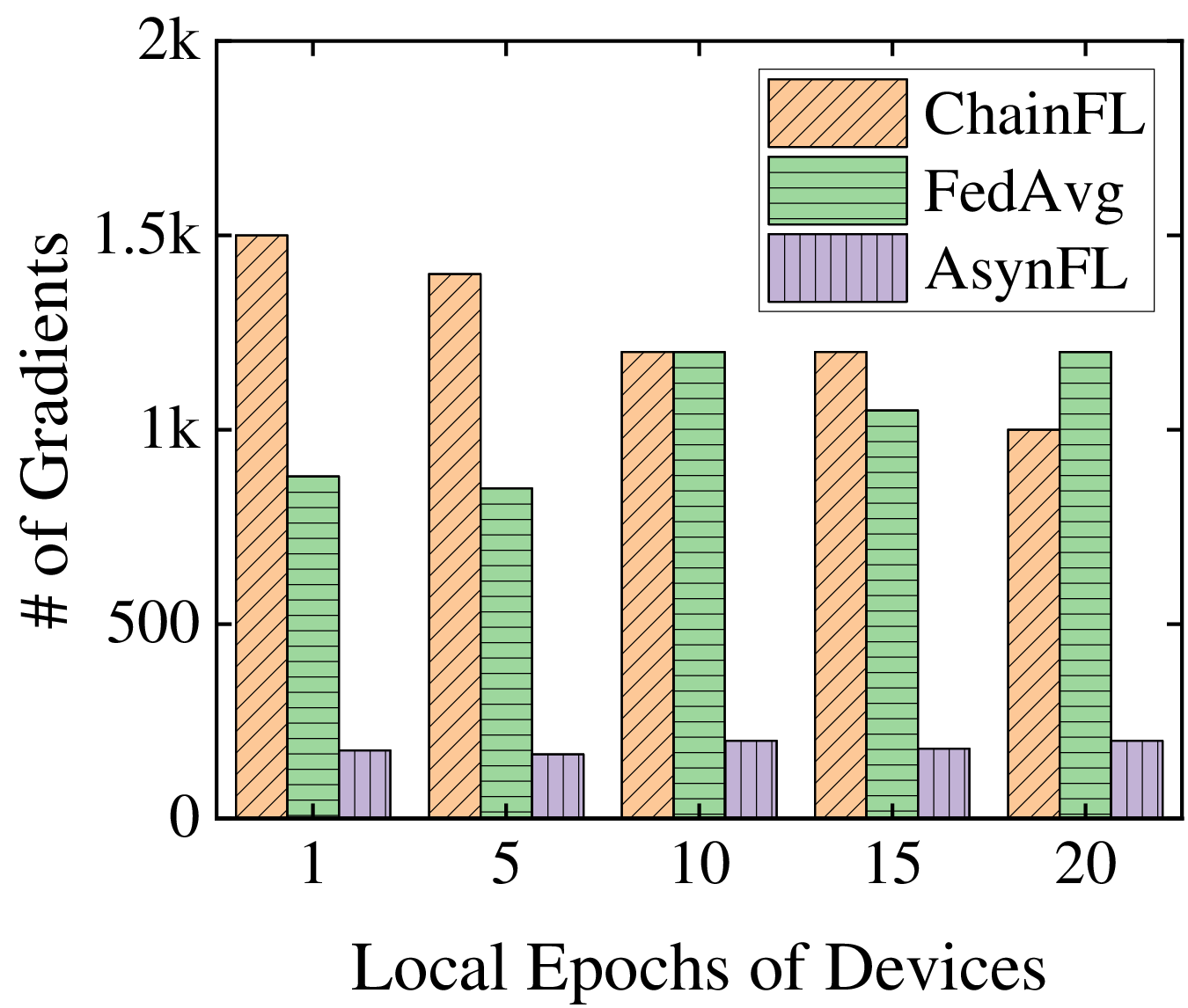}}
	\caption{Effect of the mini-batch size and local epochs of devices on \# of global epochs and \# of gradients with a preset threshold of the testing perplexity of 150 ($S_d=10$, $R=1$, $M_d=0$). (a) Global Epochs \emph{vs.} Mini-Batch Size. (b) Gradients \emph{vs.} Mini-Batch Size. (c) Global Epochs \emph{vs.} Local Epochs. (d) Gradients \emph{vs.} Local Epochs.}
	\label{fig:task2EwithGeANDGr}
\end{figure}

Our study initially evaluates FL model parameter sensitivity, focusing on mini-batch size $B \in \{10, 20, 30, 40, 50\}$ and local epochs $E \in \{1, 5, 10, 15, 20\}$. 
The training is conducted over a set number of global epochs to ascertain optimal accuracy and perplexity, with an additional analysis based on gradient numbers. 
The results of Task 1 and Task 2, in terms of the best accuracy and best perplexity, are presented in Table \ref{tab:task1Sensitivity} and Table \ref{tab:task2Sensitivity}, respectively.
For benchmarks, we use a testing accuracy threshold of 0.95 and perplexity at 150, considering training complete when these metrics are met (higher/lower than 0.95/150) by the global model. 
Model convergence is examined by comparing the number of global epochs to gradient requirements, shown in Fig. \ref{fig:task1EwithGeANDGr} and Fig. \ref{fig:task2EwithGeANDGr}.

\begin{figure}[t]
	\centering
	\subfigure[]{
		\label{fig:task2PpGe}
		\includegraphics[width=0.225\textwidth]{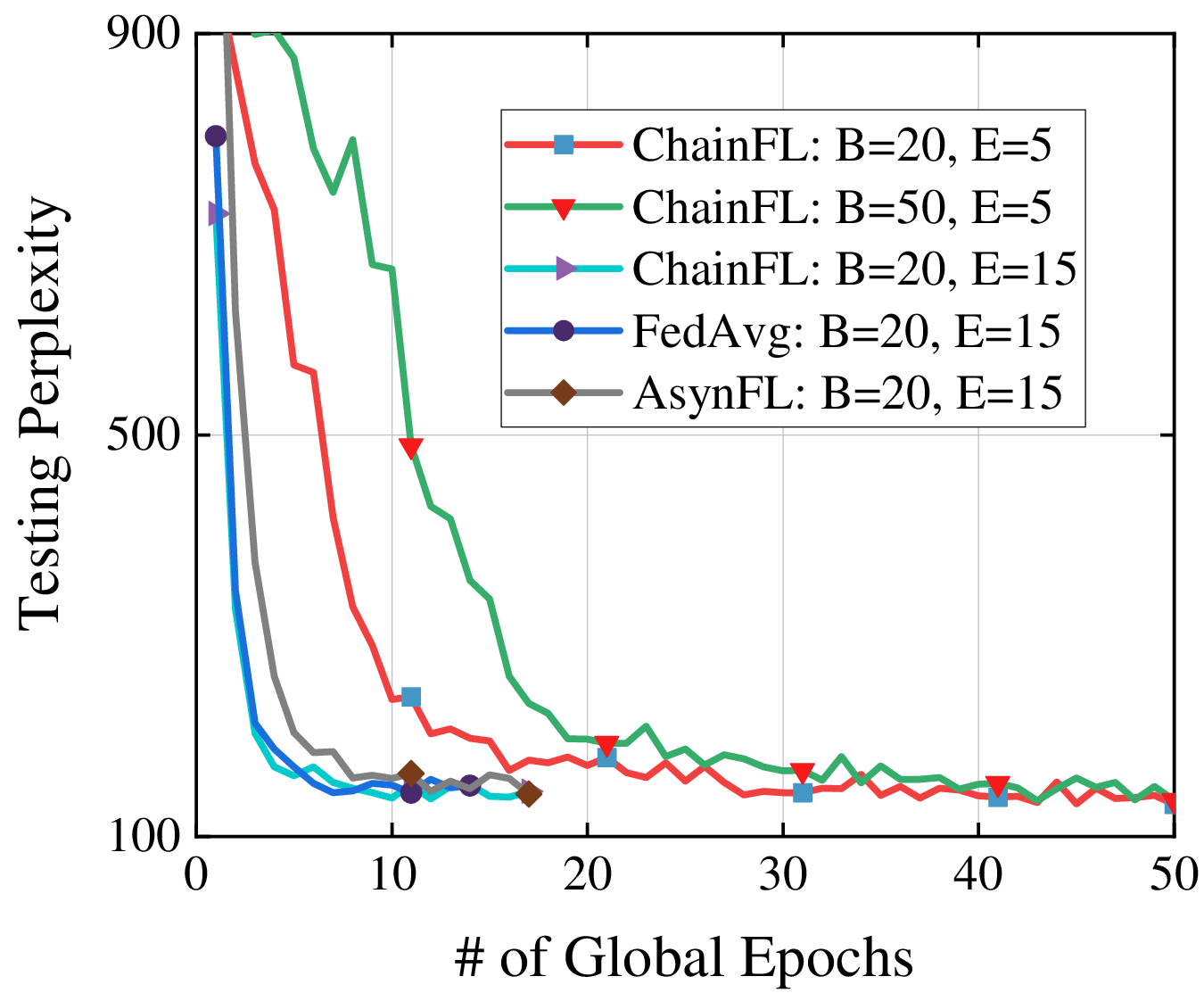}}
	\subfigure[]{
		\label{fig:task2LossGe}
		\includegraphics[width=0.225\textwidth]{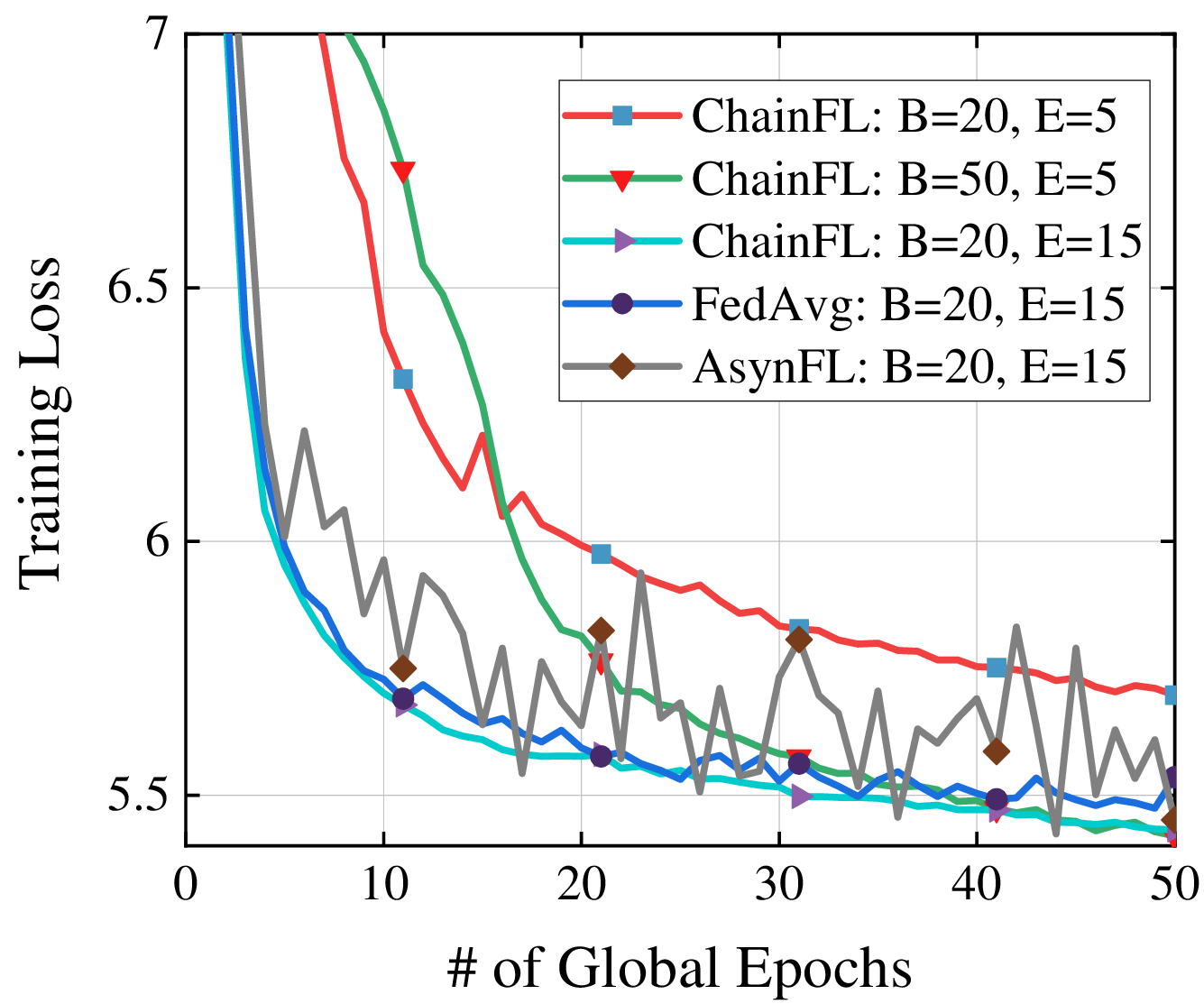}}

	\vfill

	\subfigure[]{
		\label{fig:task2PpGr}
		\includegraphics[width=0.225\textwidth]{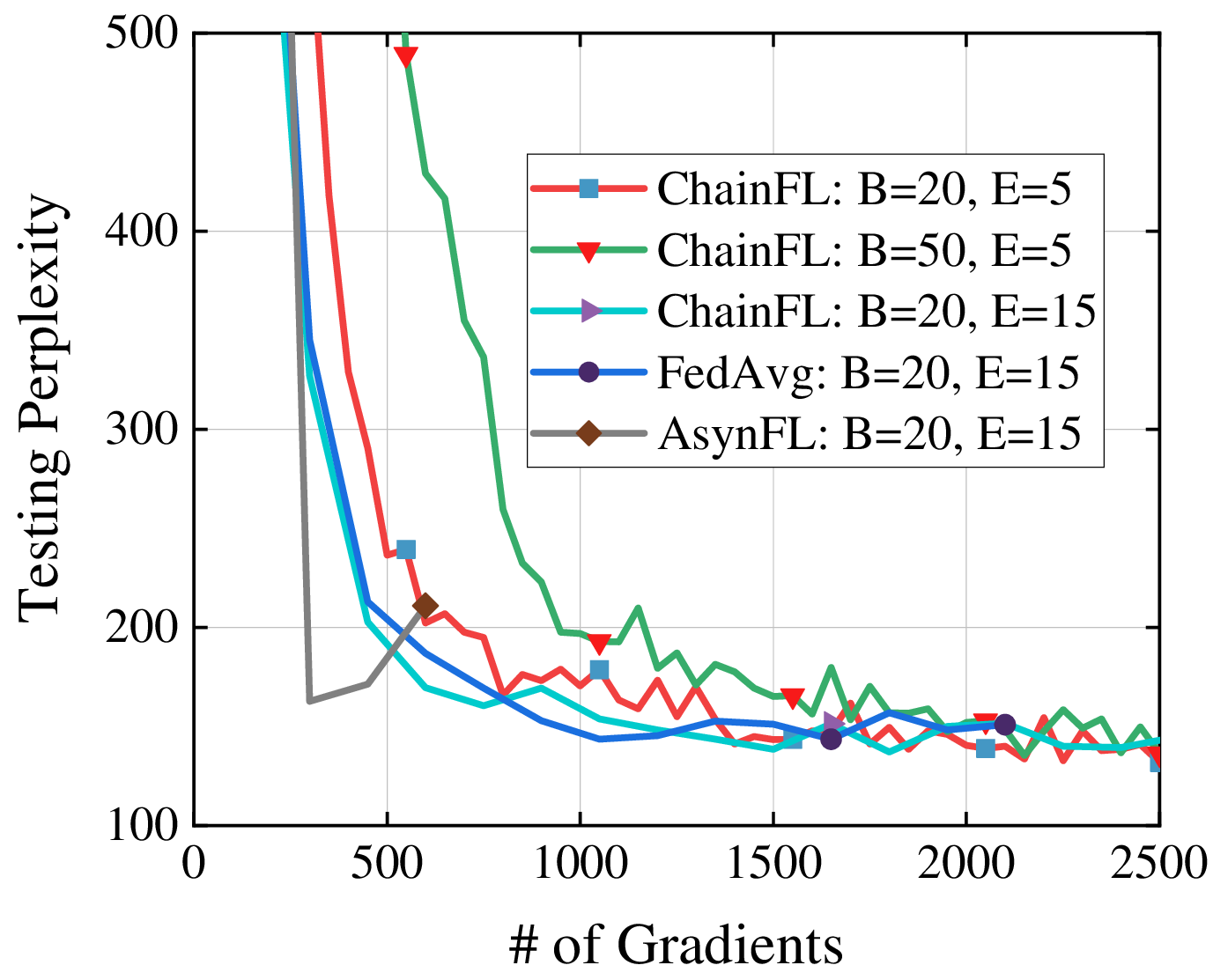}}
	\subfigure[]{
		\label{fig:task2LossGr}
		\includegraphics[width=0.225\textwidth]{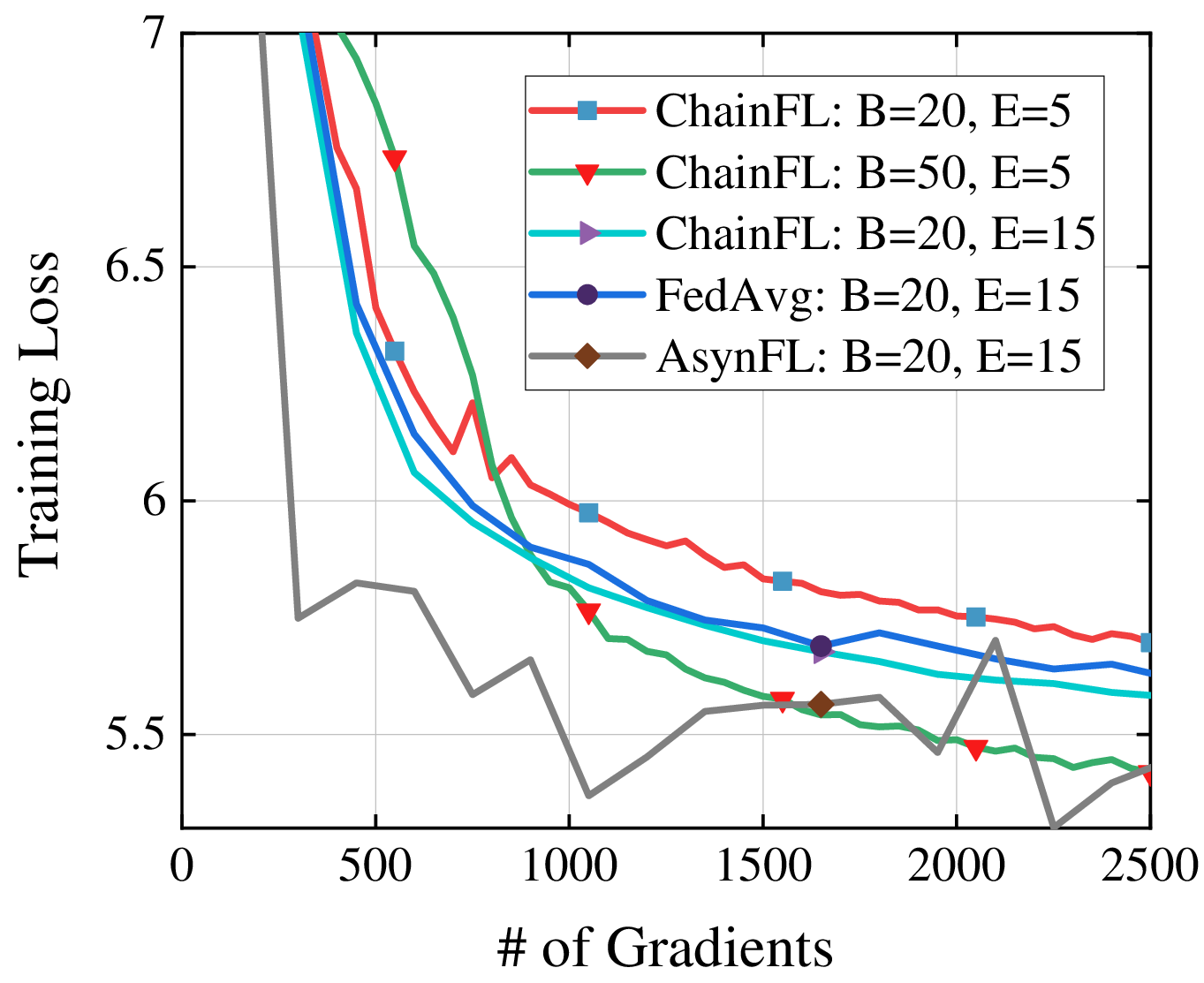}}
	\caption{Testing perplexity and training loss of Task 2 on two scales ($S_d=10$, $R=1$, $M_d=0$). (a) Perplexity \emph{vs.} Global Epochs. (b) Loss \emph{vs.} Global Epochs. (c) Perplexity \emph{vs.} Gradients. (d) Loss \emph{vs.} Gradients.}
	\label{fig:task2PpWITHLoss}
\end{figure}

\begin{figure}[t]
	\centering
	\includegraphics[width=0.47\textwidth]{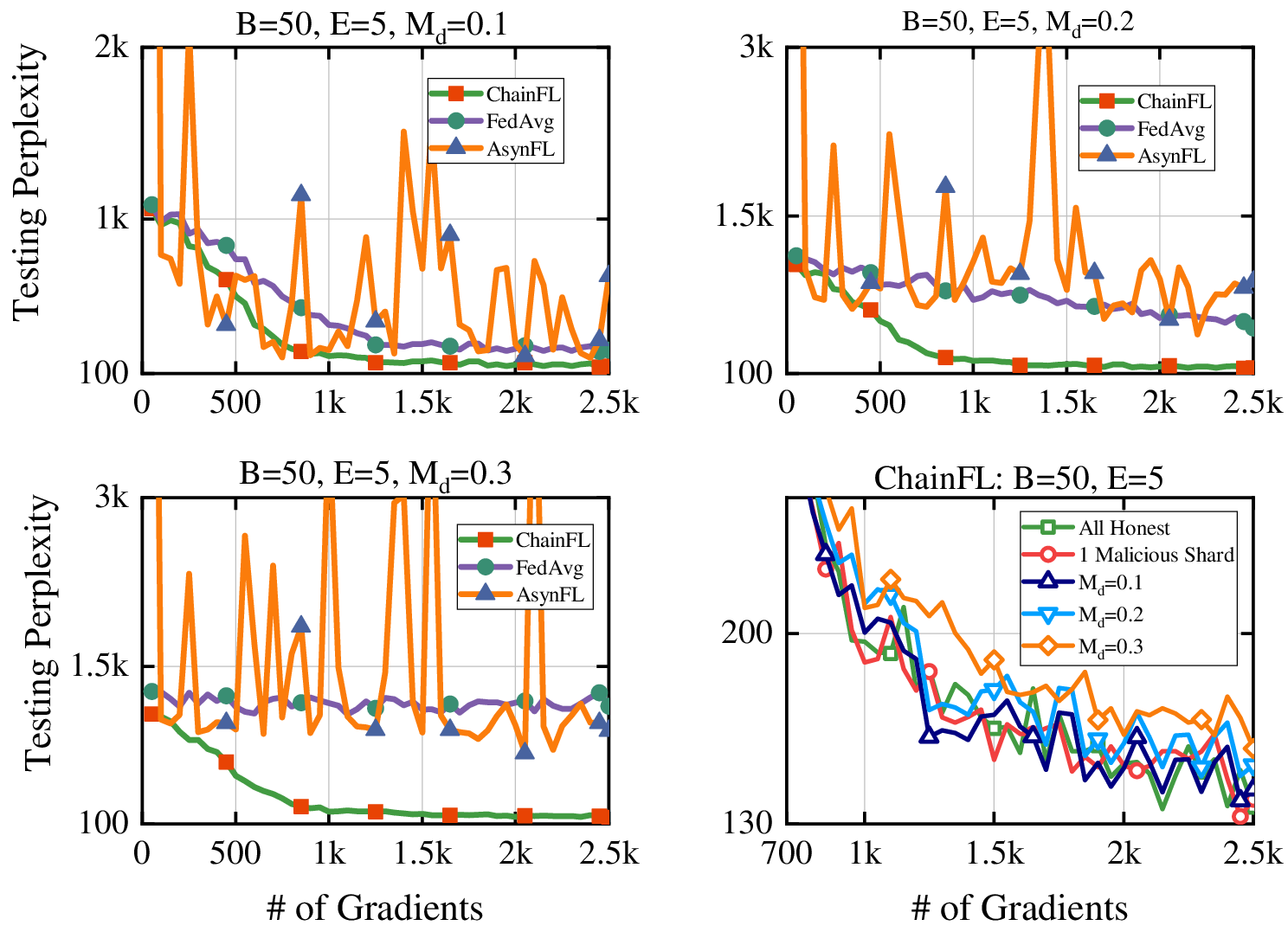}
	\caption{Effect of different malicious device ratio on the testing perplexity ($S_d=10$, $R=1$).}
	\label{fig:task2MN}
\end{figure}

\begin{figure}[t]
	\centering
	\includegraphics[width=0.47\textwidth]{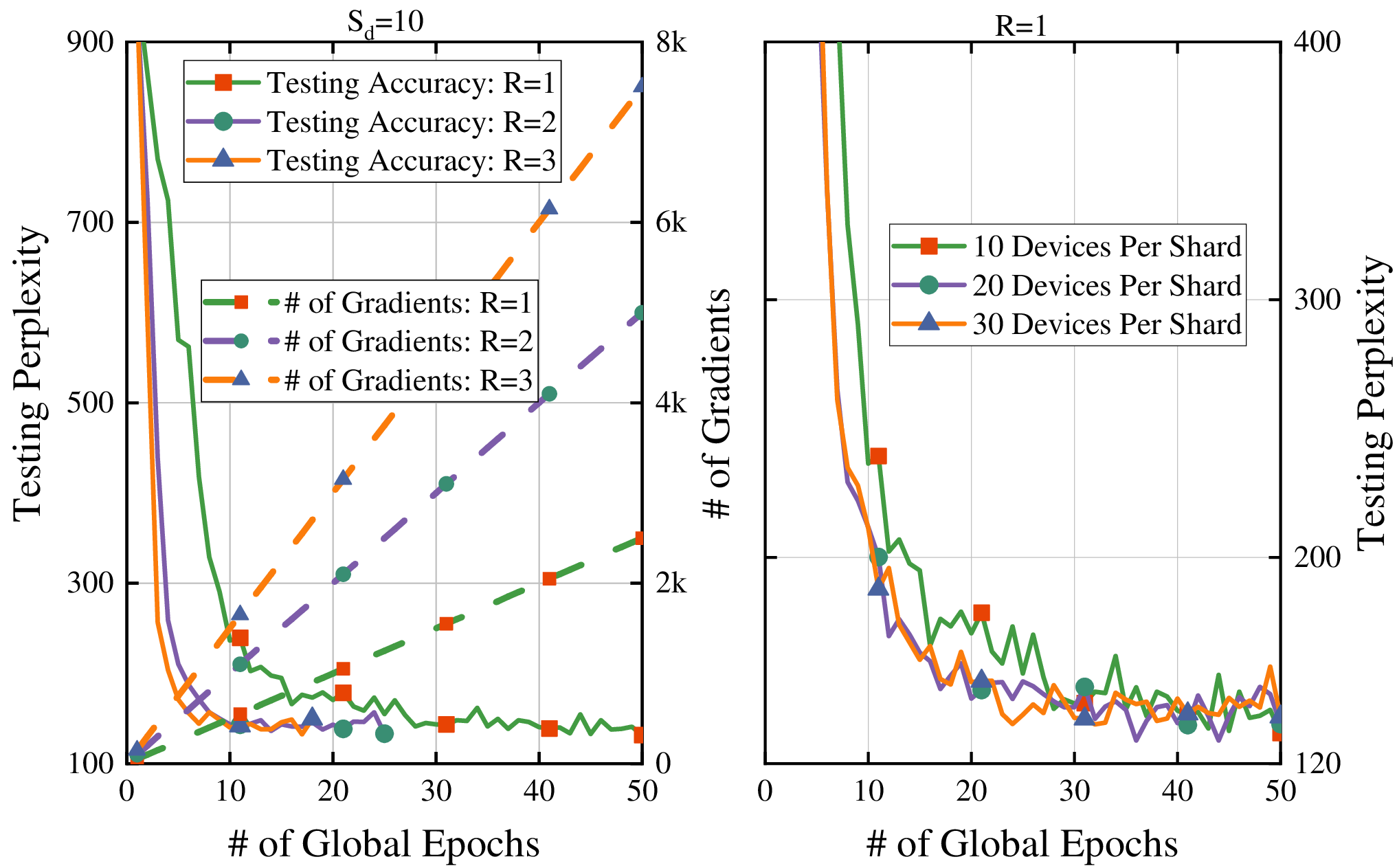}
	\caption{Effect of rounds per iteration and devices per shard on the testing perplexity ($B=20$, $R=5$, $M_d=0$).}
	\label{fig:task2roundCompare}
\end{figure}

\begin{figure}[t]
	\centering
	\includegraphics[width=0.4\textwidth]{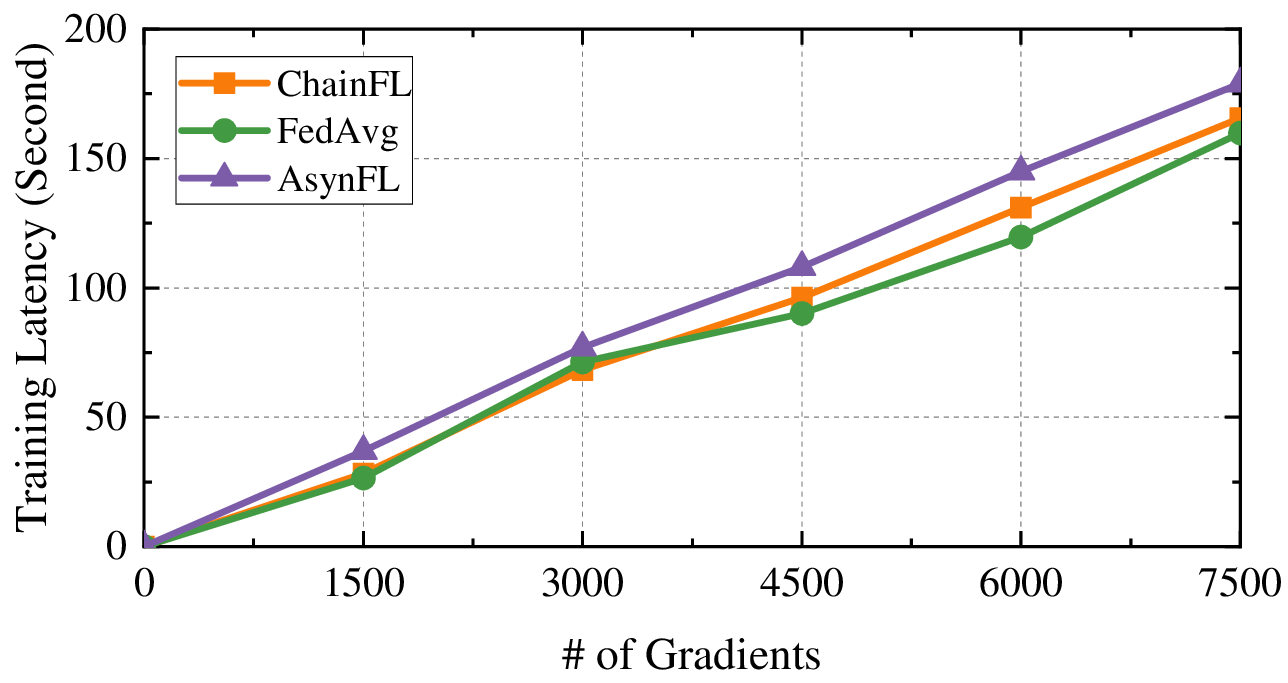}
	\caption{Training latency \emph{vs.} the number of gradients for task 1.}
	\label{fig:eva_exeTime1}
\end{figure}

\begin{figure}[t]
	\centering
	\includegraphics[width=0.4\textwidth]{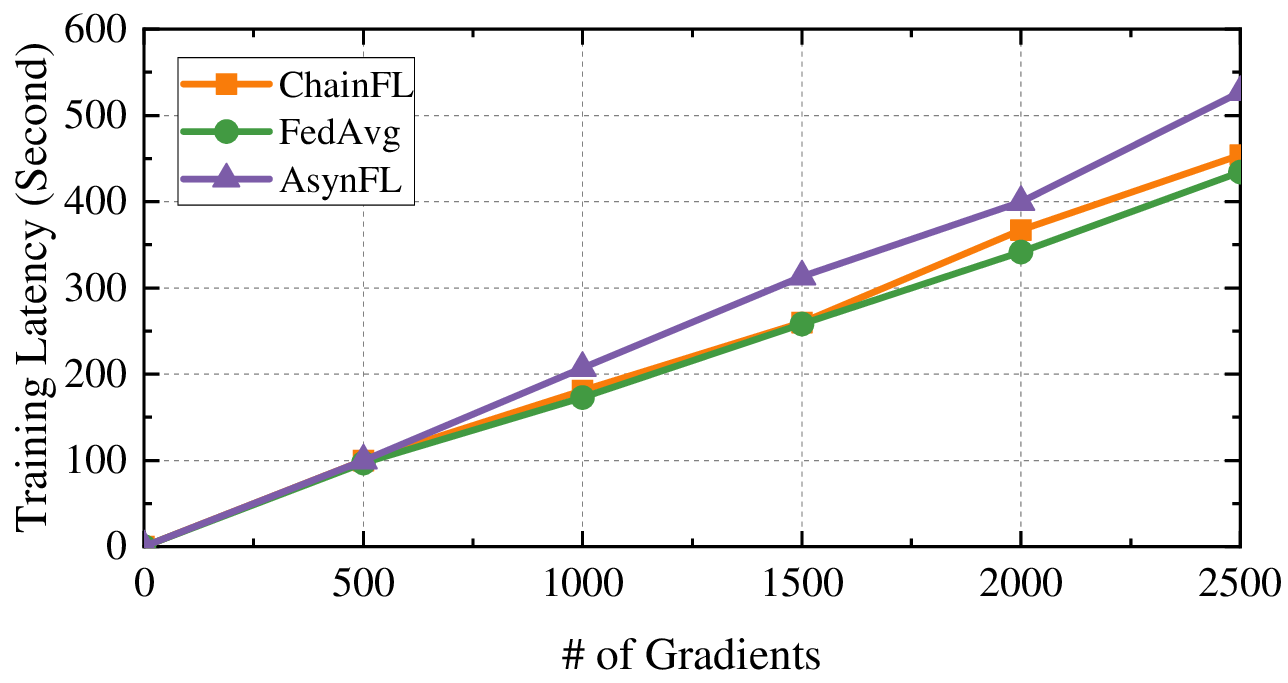}
	\caption{Training latency \emph{vs.} the number of gradients for task 2.}
	\label{fig:eva_exeTime2}
\end{figure}

Meanwhile, we present partial results of the accuracy/perplexity and loss traces with varying model parameters for FedAvg, AsynFL, and ChainFL in Fig. \ref{fig:task1AccWITHLoss} and Fig. \ref{fig:task2PpWITHLoss}.
Moreover, the resilience of these training paradigms against multiple malicious devices is evaluated, with impacts on Task 1 and Task 2 shown in Fig. \ref{fig:task1MN} and Fig. \ref{fig:task2MN}.
We also examine the influence of rounds per shard training iteration and device quantity per shard on the global model of ChainFL, presented in Fig. \ref{fig:task1roundCompare} and Fig. \ref{fig:task2roundCompare}.
Finally, the impact of integrating blockchain into FL on the training latency is evaluated, with results shown in Fig. \ref{fig:eva_exeTime1} and Fig. \ref{fig:eva_exeTime2}.

\emph{Task 1: MNIST.}
In this handwritten digit image classification task, we analyze the best accuracy over 150 global epochs and 7000 gradients produced during training.
As shown in the `Stop@ \# of Global Epochs = 150' column in Table \ref{tab:task1Sensitivity}, ChainFL outperforms FedAvg and AsynFL in global model accuracy for most mini-batch sizes and local epochs.
For instance, in the case of $B=10$ and $E=1$, ChainFL shows a roughly 14\% improvement in accuracy. 
This is credited to SLNs in ChainFL building the basic iteration model from the DAG-based mainchain, which incorporates models from all shards.
While the superiority of ChainFL is slightly reduced in the `Stop@ \# of Gradients = 7000' scenario compared to AsynFL, it still maintains an edge over FedAvg.

The influence of mini-batch size and local epochs on devices is further illustrated in Fig. \ref{fig:task1EwithGeANDGr} and Fig. \ref{fig:task1AccWITHLoss}.
For details, $E = 5$ in Fig. \ref{fig:task1Bge} and Fig. \ref{fig:task1Bgr}, and $B = 10$ in Fig. \ref{fig:task1Ege} and Fig. \ref{fig:task1Egr}.
We observe that decreasing the mini-batch size reduces the number of global epochs and gradients required to reach a testing accuracy of 0.95 in ChainFL and FedAvg.
However, increasing the amount of computation of each device by increasing local epochs does not consistently reduce global epochs, which indicates a potential decrease in computational efficiency, as concluded from Fig. \ref{fig:task1EwithGeANDGr}. 
Besides, Fig. \ref{fig:task1AccGe}, Fig. \ref{fig:task1AccGr}, Fig. \ref{fig:task1LossGe}, and Fig. \ref{fig:task1LossGr} show a faster convergence rate and higher accuracy of ChainFL compared to FedAvg and AsynFL in most scenarios. 
This enhancement results from the iterative method of ChainFL, where well-trained models from various shards in the mainchain are selected for subsequent training rounds, potentially reducing the impact of lower accuracy models.
Conversely, in FedAvg and AsynFL, both the worst and best models contribute to the global model aggregation, regardless of the performance of the models trained by each device.

In addition, this paper places a substantial emphasis on enhancing robustness during the training process.
We evaluate the performance of ChainFL against FedAvg and AsynFL under various malicious device ratios, $M_{d} \in \{0.1, 0.2, 0.3\}$. 
For instance, $M_{d} = 0.1$ indicates that $10\%$ of devices in each shard are malicious, as illustrated in Fig. \ref{fig:task1MN}. 
The results demonstrate the pronounced superiority of ChainFL in model accuracy, particularly notable at higher malicious ratios ($M_{d} = 0.3$), where ChainFL exhibits a threefold increase in robustness.
In scenarios with $M_{d} = 0.2$ and $M_d = 0.3$, the global model accuracy of FedAvg and AsynFL struggles to converge beyond 0.5 even after $7.5k$ gradients. 
In contrast, the performance of ChainFL converges to values exceeding 0.8 under the same conditions. 
Besides, the impact of varying levels of malicious activity within ChainFL is depicted in the bottom right corner of Fig. \ref{fig:task1MN}.
It is observed that, although the accuracy of ChainFL marginally diminishes with an increase in the malicious ratio, the decline is relatively minimal compared to the substantial accuracy reduction experienced by FedAvg and AsynFL.

We also compare various ChainFL parameters, including the number of rounds ($R \in \{1,2,3\}$) per shard training iteration and devices per shard, with results shown in Fig. \ref{fig:task1roundCompare}. 
The results show that a higher $R$ does not improve global model accuracy, which implies that extra computational effort is not commensurately beneficial.
Conversely, adding more devices per shard improves both the peak accuracy and the convergence speed of the global model accuracy.

\begin{table*}[t]
	\centering
	\caption{Best Perplexity of Task 2 Under Different Experimental Settings of Mini-Batch Size (B) and \# of Local Epochs (E).}
	\label{tab:task2Sensitivity}
	\begin{tabular}{clllllllllll}
		\hline
		                      &                       & \multicolumn{10}{c}{Best Perplexity}                                                                                                                                                                                                            \\
		\multicolumn{1}{l}{}  &                       & \multicolumn{5}{c}{Stop@ \# of Global Epochs=80} & \multicolumn{5}{c}{Stop@ \# of Gradients=3000}                                                                                                                                               \\
		                      & \multicolumn{1}{c}{B} & 10                                               & 20                                             & 30                & 40                & 50                & 10                & 20                & 30                & 40       & 50       \\ \hline
		\multirow{3}{*}{E=5}  & FedAvg                & 119.1973                                         & 129.0462                                       & 128.5419          & 128.6406          & 132.7023          & 119.1973          & 129.0462          & 128.5419          & 128.6406 & 132.7023 \\
		                      & AsynFL                & 132.2897                                         & 138.1707                                       & 137.5729          & 140.7736          & 147.04            & 132.2897          & 138.1707          & 135.3072          & 128.9097 & 125.5916 \\
		                      & ChainFL               & 119.302                                          & \textbf{124.0372}                              & \textbf{126.8396} & 129.7998          & \textbf{129.1252} & 119.302           & \textbf{124.0372} & \textbf{126.8396} & 134.0198 & 135.4396 \\ \hline
		\multicolumn{1}{l}{}  & \multicolumn{1}{c}{E} & 1                                                & 5                                              & 10                & 15                & 20                & 5                 & 10                & 15                & 20       &          \\ \hline
		\multirow{3}{*}{B=10} & FedAvg                & 154.8811                                         & 129.0462                                       & 133.6719          & 143.6469          & 138.19            & 129.0462          & 133.6719          & 143.6469          & 138.19   &          \\
		                      & AsynFL                & 239.7618                                         & 138.1707                                       & 141.9482          & 142.396           & 149.2502          & 138.1707          & 141.9482          & 142.396           & 149.2502 &          \\
		                      & ChainFL               & 172.6349                                         & \textbf{124.0372}                              & \textbf{131.1227} & \textbf{137.1788} & 144.9491          & \textbf{124.0372} & \textbf{131.1227} & \textbf{137.1788} & 144.9491 &          \\ \hline
	\end{tabular}
\end{table*}

\emph{Task 2: Penn Treebank.}
For Task 2, we also conduct training for a preset number of global epochs and gradients to determine the best perplexity, as shown in Table \ref{tab:task2Sensitivity}.
ChainFL consistently achieved lower global model perplexity in most cases, evident in both the `Stop@ \# of Global Epochs=80' and `Stop@ \# of Gradients=3000' columns.
In Fig. \ref{fig:task2EwithGeANDGr}, we assess the number of global epochs and gradients required to reach the target perplexity of 150.
Notably, both metrics increase with larger mini-batch sizes, while more local epochs which improve computational parallelism of devices, reduce the number of global epochs. 
However, this does not always correlate with proportional benefits in computation costs, as illustrated in Fig. \ref{fig:task2Egr}.
Fig. \ref{fig:task2PpWITHLoss} traces the testing perplexity of the global model and the training loss on two scales under the settings of $B \in \{20,50\}$ and $E \in \{5,15\}$.
ChainFL demonstrates superior performance compared with FedAvg and AsynFL in the $B=20$, $E=15$ configuration on the global epochs scale. 
Moreover, the convergence and perplexity of ChainFL also outperform FedAvg on the number of gradients scale.

Task 2 also assesses the resistance of ChainFL to malicious devices, with results displayed in Fig. \ref{fig:task2MN}. 
ChainFL exhibits greater stability in testing perplexity compared to FedAvg and AsynFL, especially as the ratio of malicious devices increases. 
The fourth subfigure in Fig. \ref{fig:task2MN} particularly highlights the robustness of ChainFL against malicious attacks. 
In addition, the impact of varying the number of rounds $R \in \{1,2,3\}$ and devices per shard $S_d\in \{10,20,30\}$ in Task 2 is examined. 
As shown in Fig. \ref{fig:task2roundCompare}, there is a slight decrease in the testing perplexity of the global model with increased $R$ and $S_d$, for the same reasons discussed in Task 1 of Fig. \ref{fig:task1roundCompare}.

Extensive experiments on Task 1 and Task 2 reveal key insights. 
ChainFL shares similar FL parameter ($B$ and $E$) sensitivities with FedAvg, where increased computational resources per device speed up convergence when the local training has not fully utilized the available local data.
However, higher $E$ values can cause overfitting and adversely affect the global model once local data is thoroughly utilized. 
Fig. \ref{fig:task1EwithGeANDGr} and Fig. \ref{fig:task2EwithGeANDGr} show that ChainFL requires more global epochs and gradients to meet preset accuracy/perplexity targets (0.95/150). 
Despite initial lag compared to FedAvg due to multi-shard model consensus, ChainFL often achieves faster convergence and higher accuracy, as shown in Fig. \ref{fig:task1AccWITHLoss} and Fig. \ref{fig:task2PpWITHLoss}.
On the other hand, in an environment without malicious devices and stale models, AsynFL exhibits fast and stable iterative updates that lead to superior performance, consistent with the findings of \cite{xie2019asynchronousfederated}.
However, the performance of AsynFL experiences a significant decline in the presence of malicious devices, as it lacks resistance against attacks.
In contrast, ChainFL demonstrates resilience against such threats, which is attributed to its local model evaluation consensus within each subchain.
Moreover, the consensus-based virtual pruning of the mainchain efficiently eliminates the malicious model published by the malicious shard to maintain a stable convergence of the accuracy of the global model.

In addition, we evaluate the impact of integrating blockchain into FL on the training latency, alongside exploring the trade-off between the latency incurred by blockchain implementation and the accuracy of FL.
To guarantee an equitable comparison, these experiments are uniformly conducted with respect to the number of gradients involved.
Fig. \ref{fig:eva_exeTime1} and Fig. \ref{fig:eva_exeTime2} illustrate the correlation between training latency and the number of gradients for task 1 and task 2, respectively.
It is discernible that the training latency for ChainFL marginally surpasses that of FedAvg. 
Nonetheless, ChainFL demonstrates superior performance over AsynFL, attributed to the accumulated latency from numerous rounds of model aggregations and queuing delays inherent in a fully asynchronous FL system.
Notably, despite the slight elevation in latency compared to FedAvg, ChainFL confers significant robustness advantages, as evidenced by up to a threefold enhancement, as depicted in Fig. \ref{fig:task1MN} and Fig. \ref{fig:task2MN}. 
In light of this trade-off, the marginal increase in latency is deemed acceptable and justifiable.

\section{Conclusions and Future Works}
\label{sec:conclusion}

In this paper, we propose ChainFL, a novel hierarchical blockchain-driven FL framework, designed to improve both the efficiency and security of FL systems in trustless edge computing environments.
We adopt a sharding architecture to parallelize the consensus among shards, thereby reducing the scale of information exchange and storage resource requirements and scaling the system throughput.
To reach a consensus on shard models, we design the cross-layer FL operation procedure and the virtual pruning of the mainchain.
Through the shard consensus and DAG-based mainchain consensus, asynchronous and synchronous optimizations are effectively combined to address the stragglers and stale models.
In addition, the hierarchical consensus enhances the robustness of ChainFL, making it more resistant to attacks from malicious entities.
The prototype of ChainFL is developed and deployed, and extensive experiments conducted on this prototype demonstrate that ChainFL provides acceptable and sometimes better training efficiency (by up to 14\%) and stronger robustness (by up to three times) compared to conventional FL systems.

For future work, we plan to investigate model replacement or double-spending attacks in ChainFL and explore an incentive mechanism based on blockchain technology to encourage IoT device participation in FL tasks.

\bibliographystyle{IEEEtran}
\bibliography{bstControlForIEEEtran,bibUsedinPaper}

\vspace{-0.4in}
\begin{IEEEbiography}[{\includegraphics[width=1in,height=1.25in,clip,keepaspectratio]{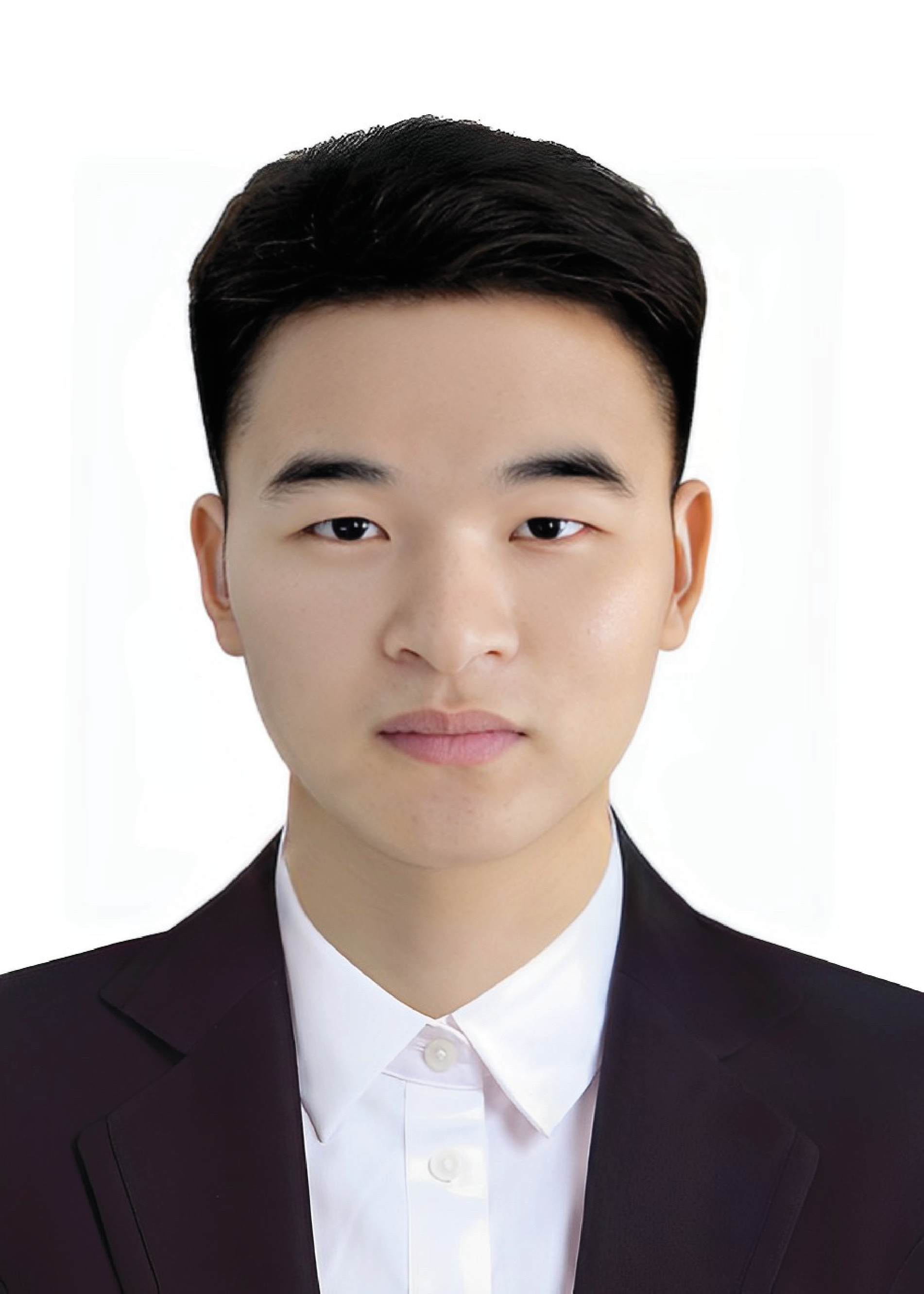}}]
	{Shuo Yuan} (Member, IEEE) received the B.S. degree from Nanchang University, Nanchang, China, and the M.E. degree in information and communication engineering from Beijing University of Posts and Telecommunications, Beijing, China, in 2016 and 2019, respectively.
	He is currently working toward a Ph.D. degree in the State Key Laboratory of Networking and Switching Technology, Beijing University of Posts and Telecommunications, Beijing, China.
	His research interests include multi-access edge computing, intelligent computing, and LEO satellite communication.
	He has been a Reviewer for IEEE \textsc{Internet of Things Journal} and IEEE \textsc{Transactions on Vehicular Technology}.
\end{IEEEbiography}
\vspace{-0.4in}

\begin{IEEEbiography}[{\includegraphics[width=1in,height=1.25in,clip,keepaspectratio]{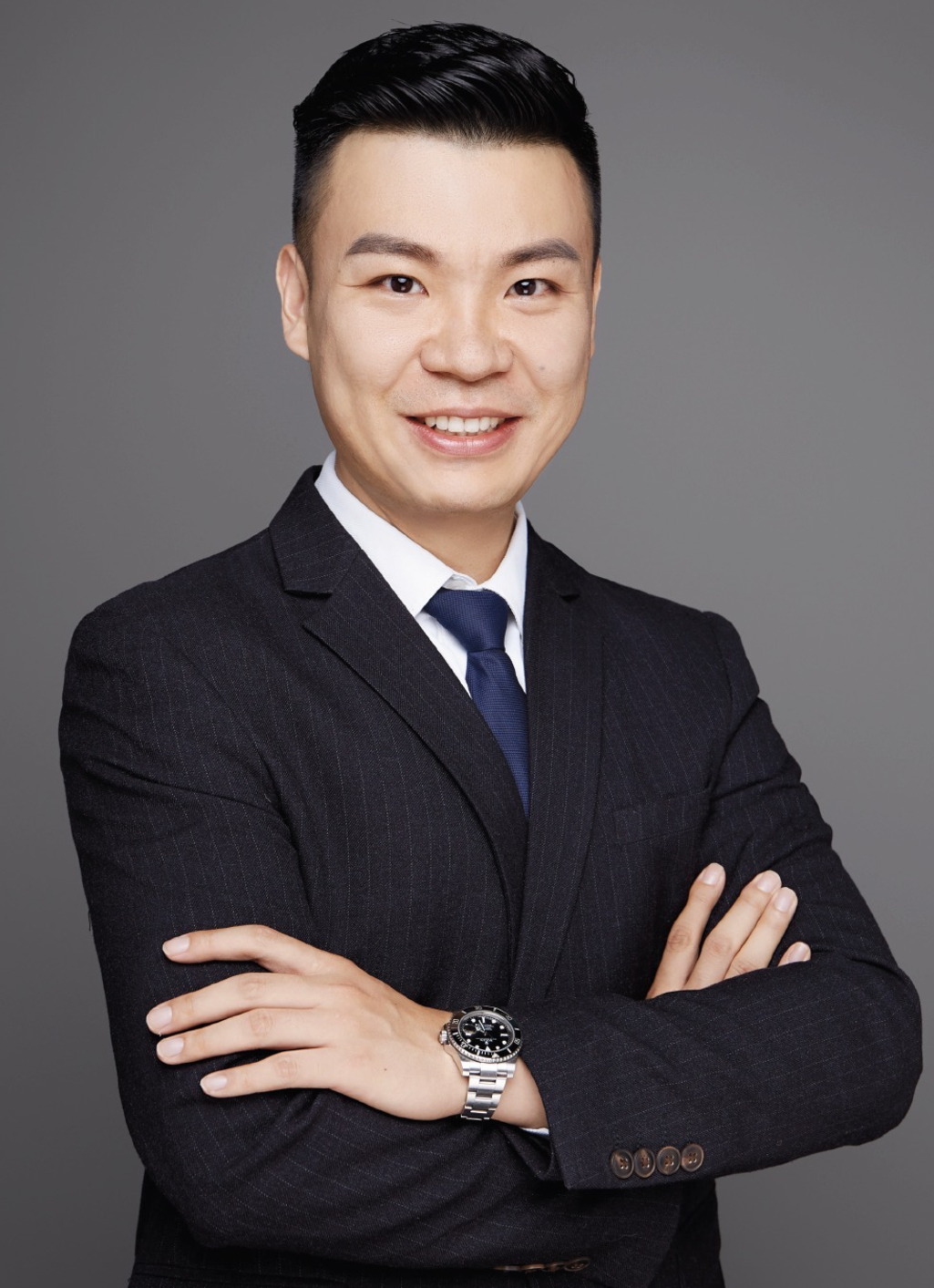}}]
	{Bin Cao} (Senior Member, IEEE) is a Professor with the State Key Laboratory of Network and Switching Technology, Beijing University of Posts and Telecommunications. 
	He received the IEEE Outstanding Leadership Award in 2020, the Best Paper Award at IEEE BMSB 2021, and the IEEE TEMS Mid-Career Award in 2021. 
	He is an Associate Editor of IEEE \textsc{Transactions on Mobile Computing} and \textsc{Digital Communications and Networks}, serves/served as a (Lead) Guest Editor of IEEE \textsc{Communications Magazine}, IEEE \textsc{Internet of Things Journal}, IEEE \textsc{Transactions on Industrial Informatics}, and IEEE \textsc{Sensors Journal}, as well as the Co-Chair for IEEE ICNC 2018, IEEE Blockchain 2020, and IEEE Globecom 2022. 
	He is the Founding Vice Chair of Special Interest Group on Wireless Blockchain Networks in IEEE Cognitive Networks Technical Committee, and a Chief Young Scientist of the National Key Research and Development Program of China.
\end{IEEEbiography}
\vspace{-0.4in}

\begin{IEEEbiography}[{\includegraphics[width=1in,height=1.25in,clip,keepaspectratio]{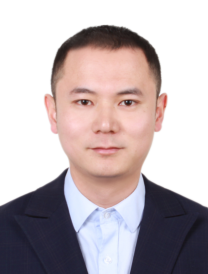}}]
	{Yao Sun} (Senior Member, IEEE) is currently a Lecturer with James Watt School of Engineering, the University of Glasgow, Glasgow, UK. 
	He has extensive research experience in wireless communication area.
	He received the IEEE Communication Society of TAOS Best Paper Award in 2019 ICC, the Best Paper Award of IEEE \textsc{Internet of Things Journal} in	2022, and Best Paper Award at IEEE ICCT in 2022.
	He has served as TPC Chair for UCET 2021, and TPC member for a number of international conferences, including ICC 2022, VTC Spring 2022, GLOBECOM 2020, WCNC 2019, ICCT 2019.
	His research interests include intelligent wireless networking, network slicing, blockchain system, internet of things and resource management in mobile networks.
\end{IEEEbiography}
\vspace{-0.4in}

\begin{IEEEbiography}[{\includegraphics[width=1in,height=1.25in,clip,keepaspectratio]{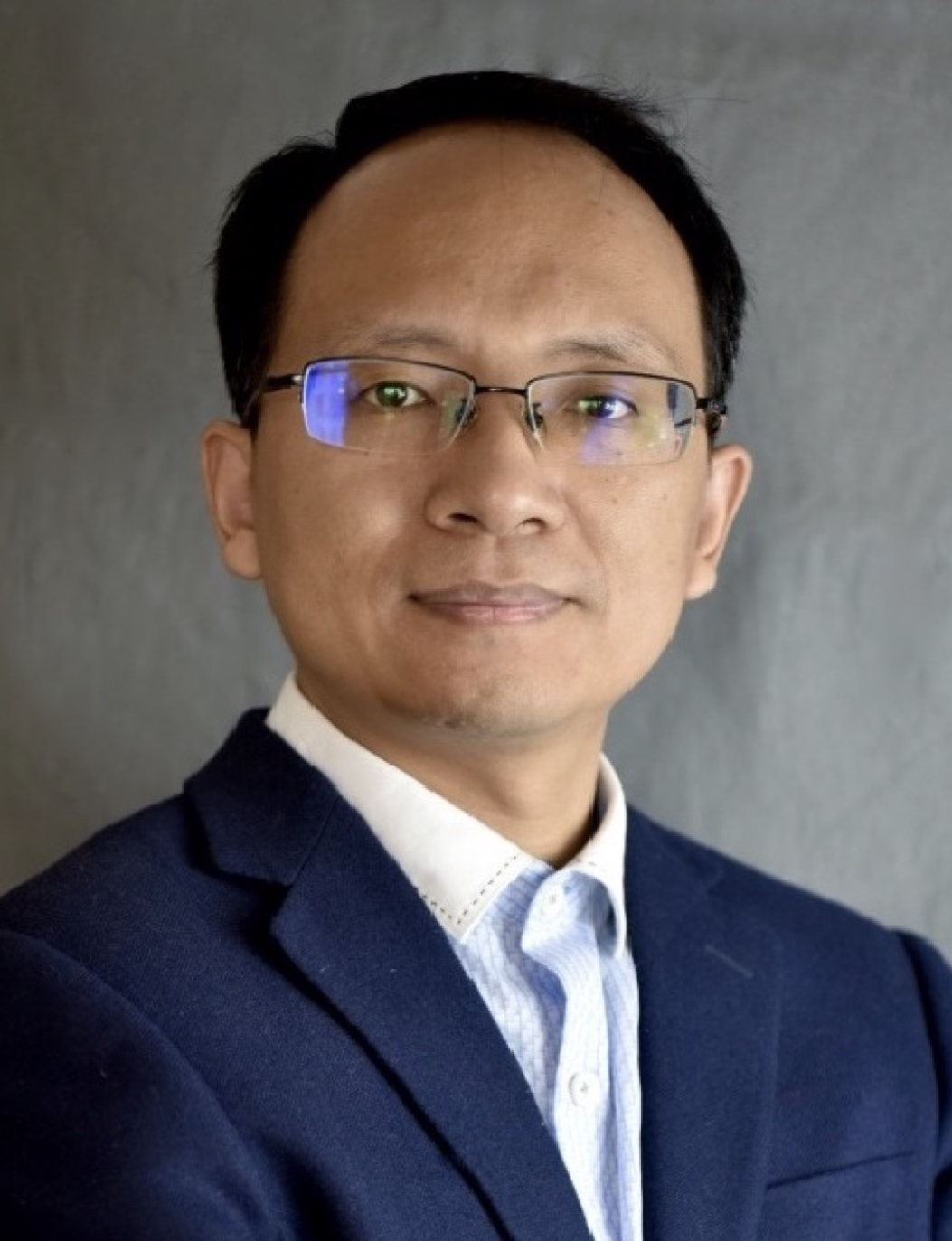}}]
	{Zhiguo Wan} (Member, IEEE) received the B.S. degree in computer science from Tsinghua University in 2002 and the Ph.D. degree from the School of Computing, National University of Singapore, Singapore, in 2007.
	He is currently a Principal Investigator with Zhejiang Laboratory, Hangzhou, Zhejiang, China.
	From 2008 to 2014, he was an Assistant Professor with the School of Software, Tsinghua University.
	He was a Post-Doctoral Researcher with the Katholieke University of Leuven, Belgium, from 2006 to 2008.
	His research interests include security and privacy for blockchain, cloud computing, and intelligent computing.
\end{IEEEbiography}
\vspace{-0.4in}

\begin{IEEEbiography}[{\includegraphics[width=1in,height=1.25in,clip,keepaspectratio]{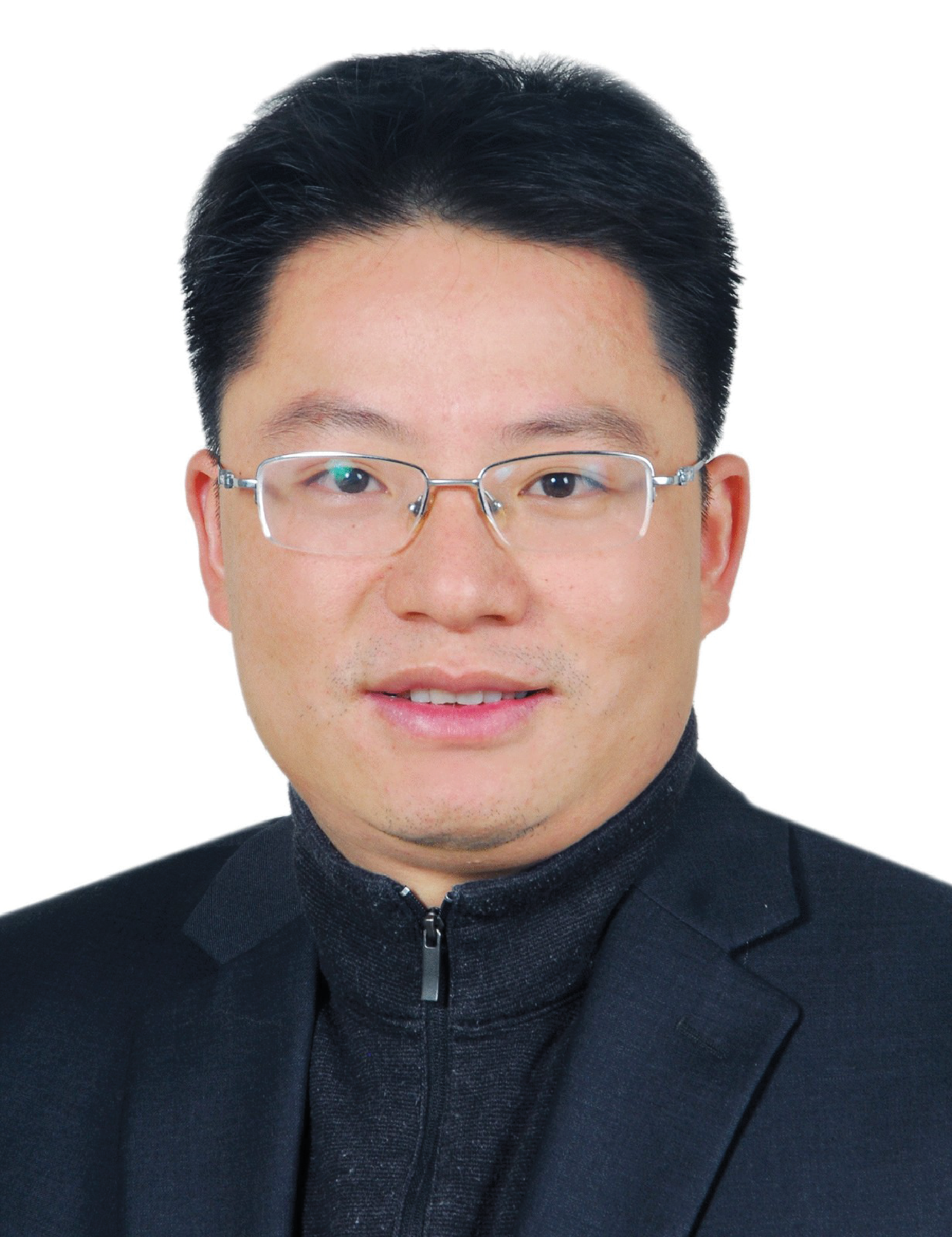}}]
	{Mugen Peng} (Fellow, IEEE) received the Ph.D. degree in communication and information systems from the Beijing University of Posts and Telecommunications, Beijing, China, in 2005. In 2014, he was an Academic Visiting Fellow at Princeton University, Princeton, NJ, USA.
	He joined BUPT, where he has been the Dean of the School of Information and Communication Engineering since June 2020, and the Deputy Director of the State Key Laboratory of Networking and Switching Technology since October 2018.
	He leads a Research Group focusing on wireless transmission and networking technologies with the State Key Laboratory of Networking and Switching Technology, BUPT.
	His main research interests include wireless communication theory, radio signal processing, cooperative communication, self-organization networking, non-terrestrial networks, and Internet of Things.
	He was a recipient of the 2018 Heinrich Hertz Prize Paper Award, the 2014 IEEE ComSoc AP Outstanding Young Researcher Award, and the Best Paper Award in IEEE ICC 2022, JCN 2016, and IEEE WCNC 2015.
	He is/was on the Editorial or Associate Editorial Board of IEEE \textsc{Communications Magazine}, IEEE \emph{Network magazine}, IEEE \textsc{Internet of Things Journal}, IEEE \textsc{Transactions on Vehicular Technology}, and IEEE \textsc{Transactions on Network Science and Engineering}.
\end{IEEEbiography}

\end{document}